\documentclass[aps,onecolumn,superscriptaddress]{revtex4}

\usepackage{graphicx}
\usepackage{amsmath}
\usepackage{amssymb}
\usepackage{xcolor}

\newcommand{\br}{\mathbf{r}}
\newcommand{\bp}{\mathbf{p}}

\begin{document}
\title{Triggering recollisions with XUV pulses: Imprint of recolliding periodic orbits}

\author{J. Dubois}
\affiliation{Aix Marseille Univ, CNRS, Centrale Marseille, I2M, Marseille, France}
\affiliation{Max Planck Institute for the Physics of Complex Systems, Dresden, Germany}
\author{\`{A}. Jorba}
\affiliation{Departament de Matem\`{a}tiques i Inform\`{a}tica, Universitat de Barcelona, Barcelona, Spain}

\begin{abstract}
We consider an electron in an atom driven by an infrared (IR) elliptically polarized laser field after its ionization by an ultrashort extreme ultraviolet (XUV) pulse. We find that, regardless of the atom species and the laser ellipticity, there exists XUV parameters for which the electron returns to its parent ion after ionizing, i.e., undergoes a recollision. This shows that XUV pulses trigger efficiently recollisions in atoms regardless of the ellipticity of the IR field.
The XUV parameters for which the electron undergoes a recollision are obtained by studying the location of recolliding periodic orbits (RPOs) in phase space. The RPOs and their linear stability are followed and analyzed as a function of the intensity and ellipticity of the IR field. 
We determine the relation between the RPOs identified here and the ones found in the literature and used to interpret other types of highly nonlinear phenomena for low elliptically and circularly polarized IR fields.
\end{abstract}

\maketitle

\section{Introduction}
Subjecting atoms to strong infrared (IR) laser fields gives rise to a variety of highly nonlinear and nonperturbative phenomena such as above-threshold ionization~\cite{Becker2002} (ATI), nonsequential multiple ionization~\cite{Huillier1983,Becker2008_ContP} (NSMI) and high harmonic generation~\cite{Huillier1991} (HHG). These highly nonlinear phenomena have tremendous applications in atomic, molecular and optical physics. For instance, they are employed to probe electron dynamics inside atoms and molecules~\cite{Lein2007,Meckel2008}, to probe electron-electron correlations~\cite{Bergues2012,Becker2012} or to generate ultrashort laser pulses~\cite{Paul2001}. By changing the ellipticity of the IR field, other properties of the atom and of the electron dynamics are probed, such as for instance Coulomb effects~\cite{Landsman2013} and nonadiabatic effects~\cite{Boge2013}. The classical mechanism built-in the physics of these highly nonlinear processes is called \emph{recollisions}~\cite{Corkum1993,Corkum2011}. A recollision is when~\cite{Corkum1993} (i) one electron leaves the atom, (ii) travels in the continuum driven by the IR field and then (iii) returns to its parent ion. When the electron returns to its parent ion, it can scatter and leave the atom with high energy~\cite{Paulus1994} (this leads to ATI), collide with one or multiple electrons and ionize~\cite{Becker2012,Mauger2009} (this leads to NSMI), or recombine into the ground state and produce high frequency photons~\cite{Gorlach2020} (this leads to HHG).

\begin{figure}
	\includegraphics[width=.7\textwidth]{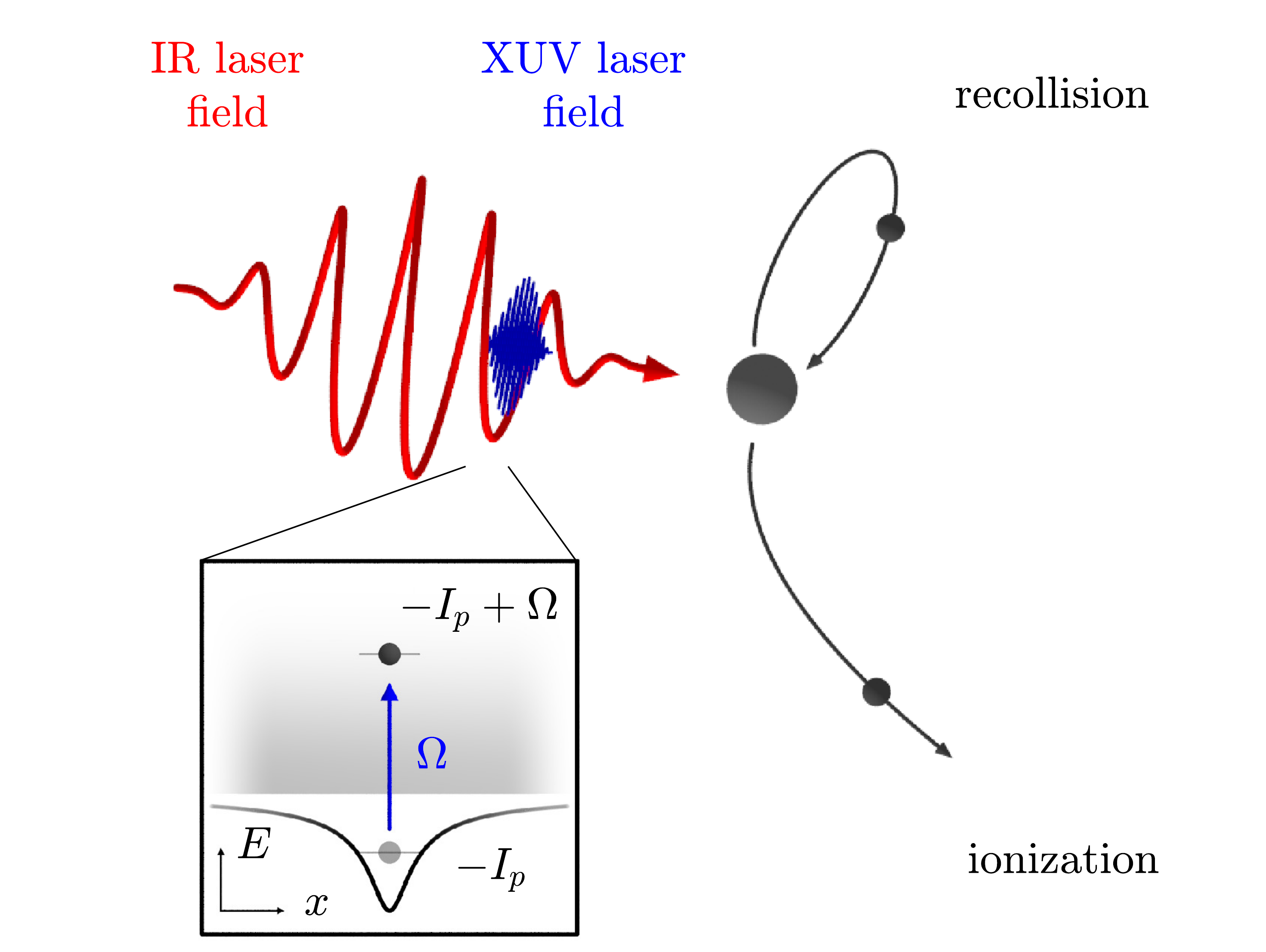}
	\caption{Schematic of pump-probe experiments~\cite{Schultze2010,Birk2020}. Initially, the electron is in the ground state of energy $- I_p$, where $I_p$ is the ionization potential. The XUV pulse of frequency $\Omega$ increases the energy of the electron from $-I_p$ to $E = -I_p + \Omega$ (see the inset panel, where the black line shows the ion-electron potential with $x$ the position of the electron). Then, the electron (small black spheres) leaves the atom (black sphere). It ionizes without recolliding (lower black curve) or returns to the core and undergoes a recollision (upper black curve).}
	\label{fig:illustration}
\end{figure}

In absence of extreme ultraviolet (XUV) pulse, i.e., in the presence of an IR field only, there are mainly two distinct ways for the electron to go outside the core region in step (i): by tunnel ionizing through the barrier induced by the strong IR field -- which is inherently a quantum process -- or by ionizing over the top of the latter barrier provided that the intensity of the IR field is large enough -- which is mainly a classical process. In the tunnel-ionization regime, which is the most commonly used in experiments, and according to tunneling theories~\cite{Ammosov1986}, ionization most probably happens around the time $t_0$ when the laser field reaches its peak amplitude. In this case, step (ii) starts at time $t_0$. At time $t_0$, the electron is near the potential barrier and its initial momentum is close to $\mathbf{p}_0 \approx \boldsymbol{0}$. The subsequent motion of the electron in step (ii) is described by classical electron trajectories~\cite{Corkum1993,Panfili2001,Becker2002}. In order to understand how the electron returns to its parent ion after ionization, we consider an IR elliptically polarized (EP) laser field $\mathbf{E} (t) = E_0 [ \mathbf{e}_x \cos (\omega t + \varphi) + \mathbf{e}_y\xi \sin (\omega t + \varphi) ] / \sqrt{\xi^2+1}$ where $\mathbf{e}_x$ and $\mathbf{e}_y$ are the major and minor polarization axes, respectively. In absence of ion-electron potential~\cite{Corkum1993}, here referred to as the strong field approximation (SFA), and in the dipole approximation, the motion of the electron is given by 
\begin{equation}
\label{eq:motion_electron_SFA}
	\mathbf{r} (t) = \mathbf{r}_0 + \left[ \mathbf{p}_0 - \mathbf{A} (t_0) \right] (t-t_0) + \dfrac{1}{\omega^2} \left[ \mathbf{E}(t) - \mathbf{E}(t_0) \right] ,
\end{equation}
where $\mathbf{r}_0 = \mathbf{r}(t_0)$ is the initial position of the electron, $\mathbf{p}_0 = \mathbf{p}(t_0)$ is its initial momentum and $\mathbf{E}(t) = - \partial \mathbf{A}(t) / \partial t$ where $\mathbf{A}(t)$ is the vector potential of the IR field. Atomic units are used unless stated otherwise. From Eq.~\eqref{eq:motion_electron_SFA}, we observe that the motion of the ionized electron is governed by two main components: A drift motion governed by the drift momentum $\mathbf{p}_0 - \mathbf{A}(t_0)$ and oscillations governed by $\mathbf{E}(t) / \omega^2$. The drift momentum also corresponds to the momentum of the averaged trajectory of the electron. The trajectory of the electron oscillates around its averaged trajectory at a frequency $\omega$ and with an amplitude $E_0 / \omega^2$. For strong IR fields (intensity $I \sim 10^{12} - 10^{16} \; \mathrm{W}\ \mathrm{cm}^{-2}$) and a laser wavelength of $780 \; \mathrm{nm}$, the excursion of the electron is $E_0 / \omega^2 \sim 1.5 - 150$ times the characteristic distance between the electron and the ionic core in the field-free atom. Due to these oscillations, the electron goes far away from the ionic core, returns towards the parent ion when the IR field changes direction and then recollides. However, if the drift momentum is too large, the electron drifts away from the core and never returns~\cite{Corkum1993}. In order to return back to the core, the drift momentum of the electron must be small compared to the characteristic velocity of an electron in an IR field (i.e., $E_0 / \omega$), which is the case if 
\begin{equation}
\label{eq:return_drift_condition}
	\mathbf{p}_0 \approx \mathbf{A} (t_0) .
\end{equation}
Typically, the minimum of the norm of the vector potential is $|\mathbf{A}(t_0)| \sim (E_0 / \omega ) \xi / \sqrt{\xi^2+1}$. 
In absence of XUV pulse, after tunnel ionizing due to the IR field~\cite{Corkum1993,Ammosov1986}, $\mathbf{p}_0 \approx \boldsymbol{0}$. Therefore, in absence of XUV pulse, Eq.~\eqref{eq:return_drift_condition} is fulfilled for low ellipticities only. As a consequence, the electron returns to its parent ion for low ellipticities only.  
We note, however, that recollisions in absence of XUV pulse can be observed for specific target atoms and laser wavelengths for IR fields with high ellipticities thanks to the combination between the variations of the laser envelope and the ion-electron interaction~\cite{Dubois2020}. Here, the laser envelope is constant. 
\par
In this article, we consider the ionization in step (i) to be induced by an ultrashort XUV pulse, as illustrated in Fig.~\ref{fig:illustration}. Such experimental setup is often referred to as a pump-probe experiment~\cite{Schultze2010,Birk2020}.
This setup has been used for different applications, for instance for enhancing HHG~\cite{Gaarde2005, Biegert2006, Gademann2011} or for probing the scattering of the electron by the parent ion~\cite{Mauritsson2008}. Here, the questions we address are: Are there XUV parameters for which recollisions occur regardless of the ellipticity of the IR field and regardless of the target atom~? Can we control and predict the XUV parameters for which recollisions are observed~?
The time $t_0$ when the electron ionizes corresponds to the time at which the XUV pulse reaches its peak amplitude. We consider that the XUV pulse is ultrashort and we neglect its influence on the electron dynamics after ionization. The frequency of the XUV pulse is $\Omega$ and its direction is $\mathbf{e}_{\rm XUV} = \mathbf{e}_x\cos \Theta + \mathbf{e}_y \sin \Theta$, which is in the polarization plane of the IR field. The angle $\Theta$ is referred to as the XUV angle and is one of the parameters varied in this study. According to first-order quantum perturbation theory, the most probable initial conditions of the electron after ionization by the XUV pulse~\cite{Saalmann2020} are given by $\mathbf{r}_0 \approx \boldsymbol{0}$ and
\begin{equation}
\label{eq:initial_momentum_XUV}
	\mathbf{p}_0 \approx \pm \mathbf{e}_{\rm XUV} \sqrt{2 \left[ \Omega - I_p - V (\mathbf{r}_0) \right]} ,
\end{equation}
where $V (\mathbf{r})$ is the ion-electron interaction potential and $I_p$ is the ionization potential of the atom. In this way, the XUV parameters provide the initial conditions of the electron in phase space. At time $t_0$, the energy of the electron is $E = \Omega - I_p$. In the SFA, the XUV frequencies and angles for which the electron undergoes recollisions are obtained from Eq.~\eqref{eq:return_drift_condition} and the initial momentum~\eqref{eq:initial_momentum_XUV} with $V = 0$. These conditions are studied in Sec.~\ref{sec:conditions_SFA}. We also show that these conditions are often inaccurate and that the Coulomb interaction makes its presence known in different ways after ionization. As a consequence, more elaborated techniques should be used to analyze the dynamics of the electron in the combined strong IR laser and Coulomb fields.
\par
Various perturbative~\cite{Goreslavski2004, Dubois2019, Popruzhenko2021} and non-perturbative~\cite{Barrabes2012, Kamor2013} methods have been developed in order to include the Coulomb interactions in the trajectory analyses. In particular, the study of periodic orbits in phase space has been used to understand and predict various highly nonlinear phenomena by fully taking into account the strong laser and Coulomb interactions. In particular, recolliding periodic orbits~\cite{Kamor2013} (RPOs), which are periodic orbits with similar shape as typical recolliding trajectories, have been used in Refs.~\cite{Kamor2013, Kamor2014, Mauger2014_JPB, Norman2015, Berman2015, Abanador2017, Dubois2020_PRE}.
In this article, we use the location of RPOs in phase space to determine the XUV frequencies $\Omega$ and angles $\Theta$ under which the electron comes back to the core after ionization, i.e., undergoes a recollision. We identify and follow RPOs for a wide range of parameters of the IR field. We show that regardless of the target atom and the ellipticity of the IR field, there exist parameters of the XUV pulse for which the electron returns to the parent ion after ionization. 
In Sec.~\ref{sec:Hamiltonian_model}, we show the Hamiltonian model for the electron dynamics in the atom driven by an IR field and the recollision probabilities as a function of the ellipticity of the IR field and the XUV frequency. We identify two families of RPOs referred to as $\mathcal{O}_1$ and $\mathcal{O}_2$. We use the SFA and the location of the RPOs in phase space to predict the XUV frequencies $\Omega$ and angles $\Theta$ for which the electron returns to its parent ion after ionization. We show that the SFA fails to predict these parameters for low ellipticities of the IR field. In contrast, we show that the RPOs predict well these parameters in the whole range of ellipticities of the IR field.
In Sec.~\ref{sec:Origin_RPOs}, we show the distinct properties of $\mathcal{O}_1$ and $\mathcal{O}_2$ by studying the primary RPOs of these two families in the field-free atom, referred to as $\mathcal{O}_1$ and $\mathcal{O}_2$. Then, we study the linear stability of $\mathcal{O}_1$ and $\mathcal{O}_2$ as a function of the intensity of the IR field for linear polarization (LP). We identify secondary RPOs, namely $\mathcal{O}_1^{\leftrightarrows}$, $\mathcal{O}_2^{\leftrightarrows}$ and $\mathcal{O}_2^{\uparrow\downarrow}$. 
In Sec.~\ref{sec:elliptically_polarized}, we follow these RPOs as a function of the ellipticity of the IR field. We show that the RPOs $\mathcal{O}_2$, $\mathcal{O}_2^{\leftrightarrows}$ and $\mathcal{O}_2^{\uparrow\downarrow}$ which co-rotates with the IR field persist at high ellipticities of the IR field in the range of parameters we have investigated.
We find the relation between the RPOs of the families $\mathcal{O}_1$ and $\mathcal{O}_2$, and the ones identified in Refs.~\cite{Kamor2013, Kamor2014, Mauger2014_JPB, Norman2015, Berman2015, Abanador2017, Dubois2020_PRE} and used to asses a variety of highly nonlinear phenomena such as for instance HHG~\cite{Kamor2014, Mauger2014_JPB, Abanador2017}, NSDI~\cite{Kamor2013, Dubois2020_PRE}, ionization stabilization~\cite{Norman2015} and the persistence of Coulomb focusing~\cite{Berman2015}. 

\section{Hamiltonian model and recollision probabilities \label{sec:Hamiltonian_model}}
In this section, we introduce the Hamiltonian model for the electron dynamics in atoms driven by an IR field in the dipole approximation. We compute the recollision probabilities for different parameters of the IR field and the XUV pulse (the parameters of the XUV pulse act on the initial conditions of the trajectories while the parameters of the IR field act on their subsequent dynamics). This provides the XUV parameters to be used, for instance, in experiments to trigger recollisions with an elliptically polarized IR field. We use the SFA to derive conditions under which the electron returns to its parent ion after ionization. We identify RPOs and use their location in phase space to obtain accurately the XUV frequencies $\Omega$ and XUV angles $\Theta$ for which the electron returns to its parent ion.

\subsection{Hamiltonian model}
We consider a $d$-dimensional configuration space. The position of the electron is denoted $\mathbf{r}$, and its canonically conjugate momentum is denoted $\mathbf{p}$. In the single-active electron approximation and the dipole approximation, the Hamiltonian governing the dynamics of the electron is
\begin{equation}
\label{eq:main_hamiltonian}
H (\mathbf{r},\mathbf{p},t) = \dfrac{|\mathbf{p}|^2}{2} + V(\mathbf{r}) + \mathbf{r} \cdot \mathbf{E} (t) .
\end{equation}
The trajectories of Hamiltonian~\eqref{eq:main_hamiltonian} are integrated using the Taylor package~\cite{Jorba2005}. 
We use the soft Coulomb potential~\cite{Javanainen1988} $V(\mathbf{r}) = - (|\mathbf{r}|^2+1)^{-1/2}$ for the ion-electron interaction, which is invariant under rotations. The IR field is elliptically polarized and reads
\begin{equation}
\label{eq:laser_field}
\mathbf{E} (t) = \dfrac{E_0}{\sqrt{\xi^2+1}} \left[ \mathbf{e}_x \cos (\omega t + \varphi ) + \mathbf{e}_y \xi \sin (\omega t + \varphi) \right] .
\end{equation}
The amplitude of the laser field is $E_0$ and is related to its intensity $I = E_0^2$. We recall that in SI units, the intensity is $I \; [\mathrm{W}\ \mathrm{cm}^{-2}] = 3.5 \times 10^{16} I \; [\mathrm{a.u.}]$. The frequency of the IR field is $\omega$, its period is $T = 2\pi/\omega$, its ellipticity is $\xi$ and the carrier-envelope phase (CEP) is $\varphi$. Here, $\varphi$ is used to model the influence of the time delay between the XUV pulse and the IR field. Throughout the article, we consider a laser wavelength of $780 \; \mathrm{nm}$, corresponding to $\omega = 0.0584 \; \mathrm{a.u.}$
The symmetries of Hamiltonian~\eqref{eq:main_hamiltonian} are given in Appendix~\ref{app:symmetries}.
 
\subsection{Recollision probabilities}

\begin{figure}
	\centering
	\includegraphics[width=\textwidth]{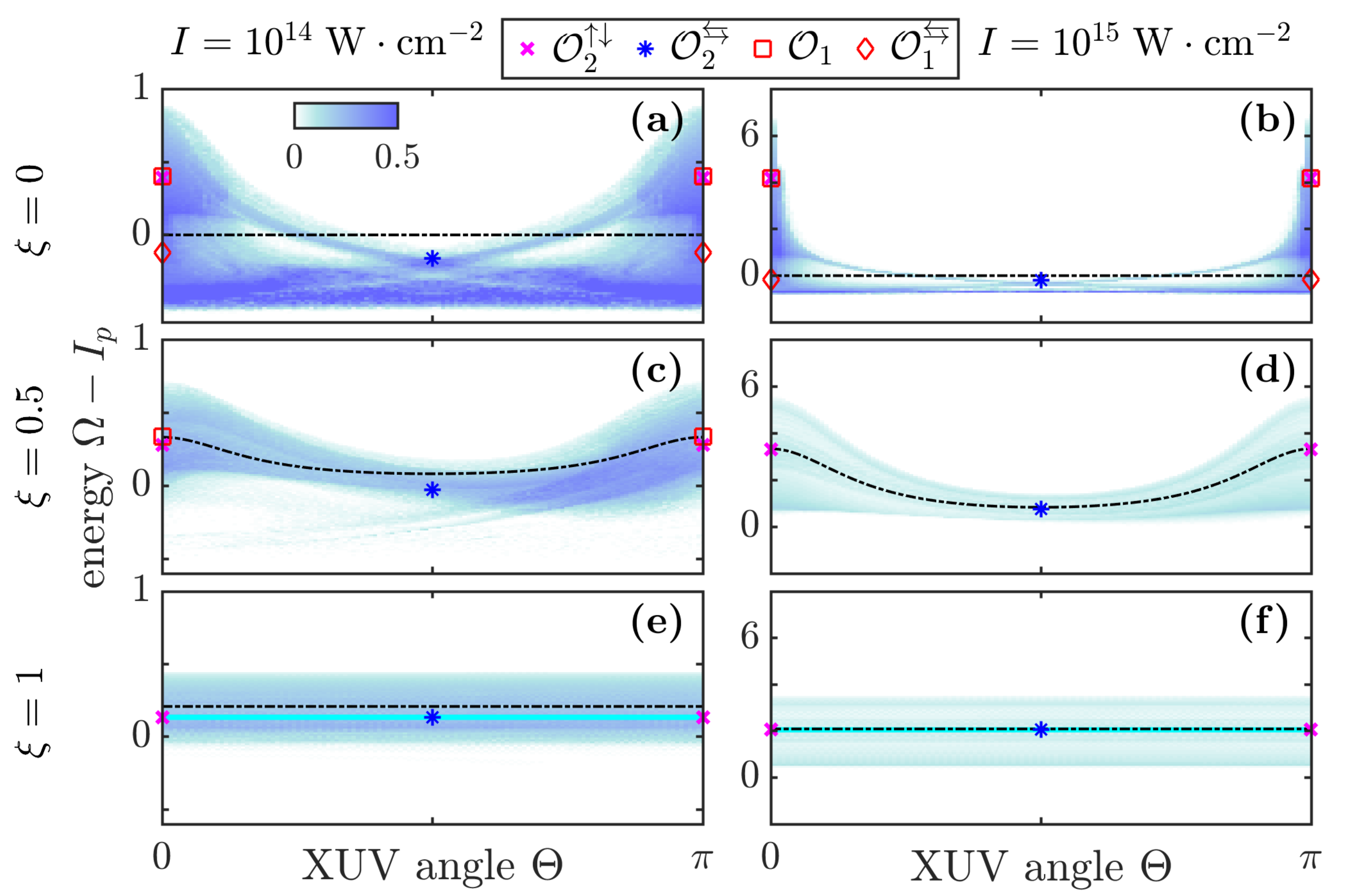}
	\caption{Recollision probabilities of Hamiltonian~\eqref{eq:main_hamiltonian} averaged over the CEP $\varphi$ as a function of the initial energy of the electron $E = \Omega - I_p$ and the XUV angles $\Theta$, for an integration time of $24 T$. The dash-dotted black curves correspond to initial energies and XUV angle which fulfill the recollision conditions in the SFA~\eqref{eq:recollision_condition_SFA}. The magenta crosses, blue asterisks, red squares and red diamonds indicate the energy and the angle of the momentum with the $\mathbf{e}_x$-axis of the RPOs $\mathcal{O}_2^{\uparrow\downarrow}$ (co-rotating with the IR field), $\mathcal{O}_2^{\leftrightarrows}$ (co-rotating with the IR field), $\mathcal{O}_1$ and $\mathcal{O}_1^{\leftrightarrows}$ when they are the closest from the origin, respectively. In the lower panels, the solid cyan curves indicate the location of the RPOs symmetric to $\mathcal{O}_2^{\uparrow\downarrow}$, $\mathcal{O}_2^{\leftrightarrows}$ with respect to the transformation~\eqref{eq:symmetry4} (see left panel of Fig.~\ref{fig:RPOs_LF_RF} and Sec.~\ref{sec:RF}) when it is the closest from the origin. On each panel, the abscissa of the markers are the angles of the momentum of the RPOs and the ordinates are their energies given by Hamiltonian~\eqref{eq:main_hamiltonian}. Energies are in a.u.}
	\label{fig:recollision_probability_theta}
\end{figure}

Figure~\ref{fig:recollision_probability_theta} shows the recollision probabilities of Hamiltonian~\eqref{eq:main_hamiltonian} averaged over the CEP $\varphi$ as a function of the initial energy of the electron $E = \Omega - I_p$ and the XUV angle $\Theta$ for $\xi = 0, 0.5$ and $1$. It is an analysis of the initial conditions leading to recollisions. The initial position of the electron is $\mathbf{r}_0 = \boldsymbol{0}$ and its initial momentum is given by Eq.~\eqref{eq:initial_momentum_XUV}. The recollision probability is the ratio between the number of recolliding trajectories and the total number of trajectories. For each pixel of Figs.~\ref{fig:recollision_probability_theta} and~\ref{fig:recollision_probability}, $128$ trajectories have been computed, each one of them corresponding to a different CEP $\varphi$ linearly spaced in the interval $[0,2\pi]$.
A trajectory of Hamiltonian~\eqref{eq:main_hamiltonian} is counted as a recollision when there exists $t_1$ and $t_2$ such that $| \mathbf{r} (t_1) | > R$ and $|\mathbf{r}(t_2) | < R$ with $t_1 < t_2$. In our calculations, $R = 5$. 
For LP fields ($\xi = 0$), we note that for $\Theta = 0$ and $\pi$, the electron dynamics is in the invariant subspace, and as a consequence its dynamics is 1D ($d=1$). Otherwise, the dynamics is inherently 2D ($d=2$). 
In Fig.~\ref{fig:recollision_probability_theta}, we observe that for all target species (i.e., for all $I_p$) and for $\xi = 0$, $0.5$ and $1$, there exist XUV frequencies $\Omega$ for which the electron undergoes recollisions. This suggests that XUV pulses can be used to efficiently trigger recollisions in atoms regardless of the ellipticity of the IR field. Also, we observe that there exist XUV frequencies $\Omega$ for which there are no recollisions. Here, we determine and compare two methods to obtain the XUV frequencies and angles for which the electron undergoes recollisions: The SFA and the location of RPOs in phase space.

\subsubsection{Recollision conditions in the SFA \label{sec:conditions_SFA}}
In the SFA (for $V = 0$), the conditions under which the electron undergoes recollisions are approximately given by Eq.~\eqref{eq:return_drift_condition}. After ionization, if the drift momentum of the electron is very small, the electron cannot drift away from the atom and can come back with the oscillations of the IR field. For the IR field given by Eq.~\eqref{eq:laser_field}, condition~\eqref{eq:return_drift_condition} is fulfilled for
\begin{subequations}
\label{eq:recollision_condition_SFA}
\begin{eqnarray}
	&& \Omega - I_p = \dfrac{\xi^2 E_0^2}{2\omega^2 (\xi^2+1)} \left( \sin^2 \Theta + \xi^2 \cos^2 \Theta \right)^{-1} , \label{eq:recollision_frequency_condition} \\
	&& \tan (\omega t_0 + \varphi) \, \tan \Theta = - \xi . \label{eq:recollision_theta_condition}
\end{eqnarray}
\end{subequations}
For LP fields ($\xi = 0$), the XUV pulse must be aligned along the laser field, i.e., $\Theta= n \pi$ with $n \in \mathbb{N}$, or the XUV frequency must be tuned such that $\Omega = I_p$. When condition~\eqref{eq:recollision_frequency_condition} is fulfilled and the IR field is CEP-unstable (i.e., if $\varphi$ is averaged over $[0,2\pi]$), there always exists $\varphi$ which fulfills Eq.~\eqref{eq:recollision_theta_condition}, regardless of the ionization time $t_0$. In contrast, when condition~\eqref{eq:recollision_frequency_condition} is fulfilled and the IR field is CEP-stable (i.e., if $\varphi$ is fixed), the ionization time $t_0$ (and hence the time when the XUV pulse reaches its peak amplitude) must be tuned in order to fulfill condition~\eqref{eq:recollision_theta_condition}. 
In Fig.~\ref{fig:recollision_probability_theta}, the dash-dotted black curves are the XUV frequencies $\Omega$ and angles $\Theta$ which fulfill Eqs.~\eqref{eq:recollision_condition_SFA}.
For $\xi = 0.5$ and $1$, we observe that Eqs.~\eqref{eq:recollision_condition_SFA} predict qualitatively well the XUV parameters for which the trajectories of Hamiltonian~\eqref{eq:main_hamiltonian} recollide.
However, for LP fields (see upper panels of Fig.~\ref{fig:recollision_probability_theta}), we observe that these conditions~\eqref{eq:recollision_condition_SFA} cannot predict accurately the XUV parameters for which the electron returns to its parent ion. In particular, around $\Theta = \pi/2$, the recollision probabilities of Hamiltonian~\eqref{eq:main_hamiltonian} is maximum for $\Omega < I_p$, while Eq.~\eqref{eq:recollision_frequency_condition} always predicts $\Omega \geq I_p$.

\subsubsection{Recollision conditions and the location of RPOs in phase space \label{sec:RPOs_return_conditions}}

Alternatively to Eqs.~\eqref{eq:recollision_condition_SFA}, we use the location of RPOs in phase space to determine and predict the conditions under which the electron undergoes recollisions. We start with the simpler case in order to illustrate the method, the 1D case ($d=1$), which is when the IR field is LP ($\xi = 0$) and the XUV angle is $\Theta = 0$ or $\pi$ [see Eq.~\eqref{eq:initial_momentum_XUV}], and then we move to the 2D case ($d=2$).
\par
\begin{figure}
	\centering
	\includegraphics[width=\textwidth]{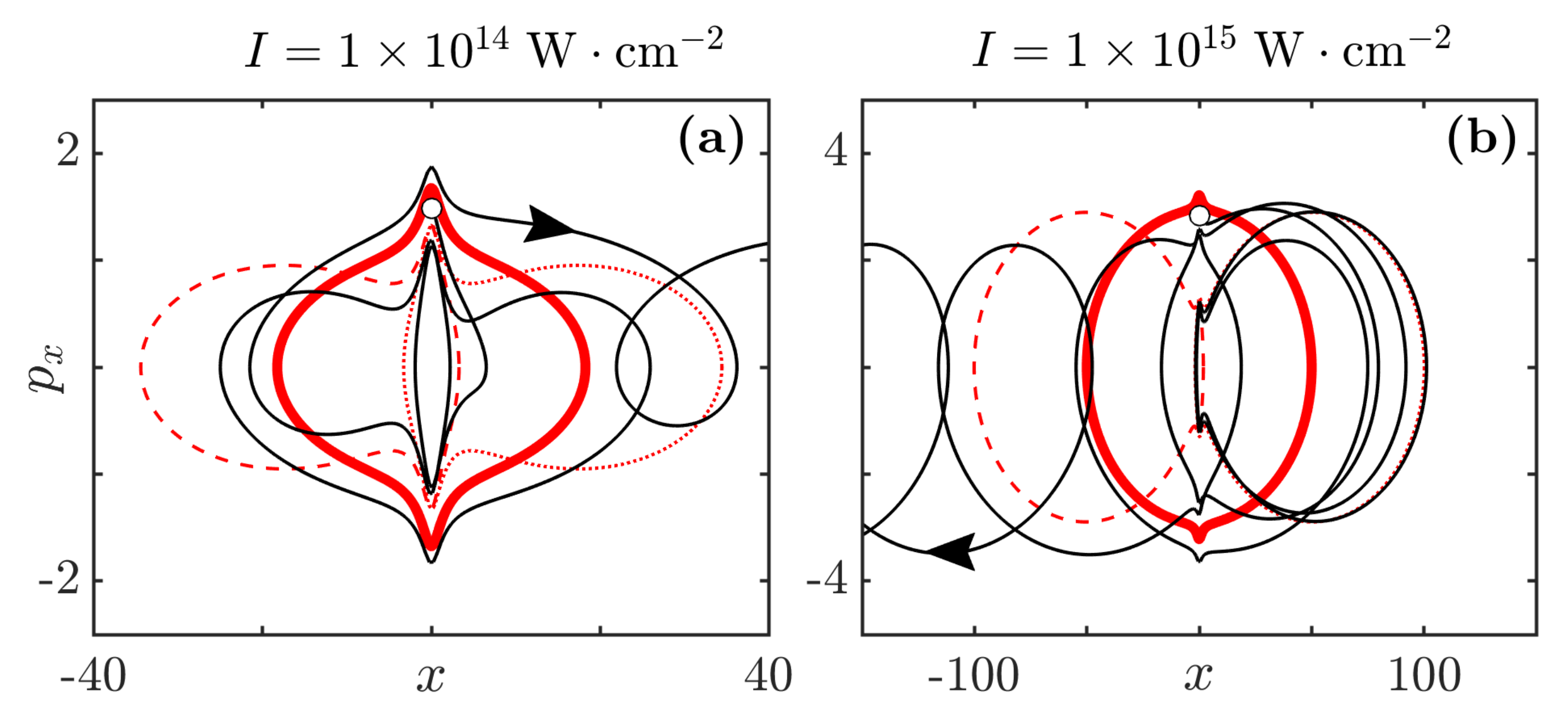}
	\caption{Typical recolliding trajectories of Hamiltonian~\eqref{eq:main_hamiltonian} (solid black curves) in phase space for $\xi = 0$ (LP fields), $d=1$ and $\Theta = 0$, for (a) $I = 10^{14} \; \mathrm{W}\ \mathrm{cm}^{-2}$ and (b) $I = 10^{15} \; \mathrm{W}\ \mathrm{cm}^{-2}$. The white markers indicate the initial conditions of the electron. The initial energy of the electron $E = \Omega - I_p$ and the CEP are, (a) $E = 0.1$ and $\varphi \approx 0.5$, and (b) $E = 3$ and $\varphi \approx 4.1$. The red solid, dashed and dotted curves are the RPOs $\mathcal{O}_1$, $\mathcal{O}_1^{\leftarrow}$ and $\mathcal{O}_1^{\rightarrow}$, respectively. Positions and momenta are in a.u.}
	\label{fig:traj1D_vs_RPOs}
\end{figure}

For 1D ($d=1$), Fig.~\ref{fig:traj1D_vs_RPOs} shows typical recolliding trajectories of Hamiltonian~\eqref{eq:main_hamiltonian} in phase space for $x_0 = 0$ and initial momentum~\eqref{eq:initial_momentum_XUV}, corresponding to recolliding trajectories in Figs.~\ref{fig:recollision_probability_theta}a and~\ref{fig:recollision_probability_theta}b for $\Theta = 0$. After ionization, we observe that the recolliding trajectories goes far away from the atom (around $E_0/\omega^2 \sim 15$ for $I= 10^{14} \; \mathrm{W}\ \mathrm{cm}^{-2}$ and $E_0/\omega^2 \sim 50$ for $I= 10^{15} \; \mathrm{W}\ \mathrm{cm}^{-2}$) and then return to their parent ion. When they return, at $x \approx 0$, their momentum is typically $E_0/\omega \sim 1$ for $I= 10^{14} \; \mathrm{W}\ \mathrm{cm}^{-2}$ and $E_0/\omega \sim 3$ for $I= 10^{15} \; \mathrm{W}\ \mathrm{cm}^{-2}$. Also, we observe a sudden peak in momentum due to the nonlinearities with the ion-electron interactions. The red solid, dashed and dotted curves are the RPOs $\mathcal{O}_1$, $\mathcal{O}_1^{\leftarrow}$ and $\mathcal{O}_1^{\rightarrow}$, respectively. We observe that these RPOs have the same shape as typical recolliding trajectories, i.e., their characteristic distance from the origin is $E_0 / \omega^2$ and their characteristic momentum is $E_0 / \omega$. Also, RPOs are periodic orbits of Hamiltonian~\eqref{eq:main_hamiltonian}, therefore they capture the effects of the ion-electron interaction. In particular, at $x \approx 0$, we observe that the RPOs capture the peak in momentum observed in the recolliding trajectories. In 1D, the particularity of the RPOs $\mathcal{O}_1$ and $\mathcal{O}_1^{\leftrightarrows}$ is that their stable and unstable manifolds act as barriers in phase space for the motion of the electron~\cite{Kamor2014, Berman2015}. As a consequence, the electron is driven by these invariant manifolds and mimic the shape of the RPOs. The location of the RPOs in phase space indicates the location of the invariant structures which drive the electron and, in particular, which drive it back to its parent ion. 
In Figs.~\ref{fig:recollision_probability_theta}a and~\ref{fig:recollision_probability_theta}b, the red squares and red diamonds indicate the energy and the angle of the momentum of $\mathcal{O}_1$ and $\mathcal{O}_1^{\leftrightarrows}$ when it crosses $x=0$ (i.e., $\Theta = 0$ or $\pi$). We observe that the red markers are in energy regions for which the recollision probability is large. In particular, we observe two main energy range for which the recollision probability is large: One for which the initial energy of the electron is negative (for $E \sim [-0.5,0]$ for $I = 10^{14} \; \mathrm{W}\ \mathrm{cm}^{-2}$ and $E \sim [-1,0]$ for $I = 10^{15} \; \mathrm{W}\ \mathrm{cm}^{-2}$) and another for which the initial energy of the electron is positive (for $E \sim [0,1]$ for $I = 10^{14} \; \mathrm{W}\ \mathrm{cm}^{-2}$ and $E \sim [0,7]$ for $I = 10^{15} \; \mathrm{W}\ \mathrm{cm}^{-2}$). We note that the negative energy range cannot be predicted by the SFA conditions~\eqref{eq:recollision_condition_SFA}. We observe that $\mathcal{O}_1$ (resp. $\mathcal{O}_1^{\leftrightarrows}$) contributes to the range for which the initial energy of the electron is positive (resp. negative). Indeed, in Fig.~\ref{fig:traj1D_vs_RPOs}, we observe that after ionization when the electron is at $x = 0$, it is closer to the RPO $\mathcal{O}_1$ (whose energy is larger) or to $\mathcal{O}_1^{\leftrightarrows}$ (whose energy is smaller) depending on the initial momentum of the electron. In both cases, the electron ionizes in the neighborhood of RPOs, which are in regions of phase space where invariant structures drive the electron back to its parent ion. If the energy of the electron after ionization is too large, there are no invariant structures which drive the electron back to their parent ion, and the electron ionizes without recolliding.

\begin{figure}
	\centering
	\includegraphics[width=\textwidth]{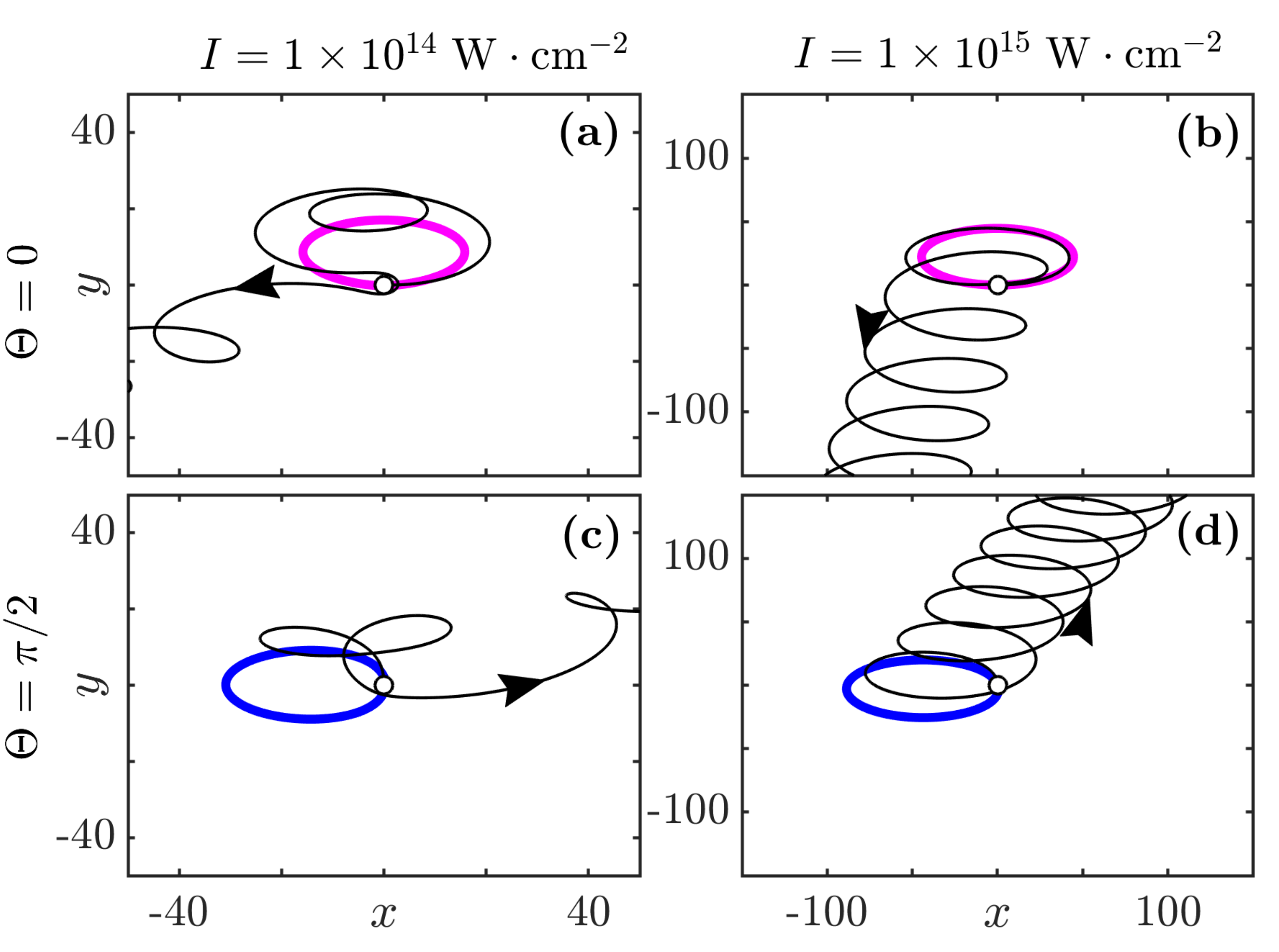}
	\caption{Typical recolliding trajectories of Hamiltonian~\eqref{eq:main_hamiltonian} (solid black curves) in the polarization plane for $\xi = 0.5$ and $d=2$, for $I = 10^{14} \; \mathrm{W}\ \mathrm{cm}^{-2}$ [left panels (a) and (c)] and $I = 10^{15} \; \mathrm{W}\ \mathrm{cm}^{-2}$ [right panels (b) and (d)], $\Theta = 0$ [upper panels (a) and (b)] and $\Theta = \pi/2$ [lower panels (c) and (d)]. The white markers indicate the initial position of the electrons $\mathbf{r}_0 = 0$. The initial energy of the electron $E = \Omega - I_p$ and the CEP are, (a) $E = 0.1$ and $\varphi \approx 4.2$, (b) $E = 3$ and $\varphi \approx 4.7$, (c) $E = 0.1$ and $\varphi \approx 6.3$, and (d) $E = 1$ and $\varphi \approx 0.06$. The magenta and blue curves are the RPOs $\mathcal{O}_2^{\uparrow}$ and $\mathcal{O}_2^{\leftarrow}$ which co-rotate with the IR field, respectively. Positions are in a.u.}
	\label{fig:traj_vs_RPOs}
\end{figure}

For 2D ($d=2$), Fig.~\ref{fig:traj_vs_RPOs} shows typical recolliding trajectories of Hamiltonian~\eqref{eq:main_hamiltonian} in phase space for $\mathbf{r}_0 = 0$ and initial momentum~\eqref{eq:initial_momentum_XUV}, corresponding to recolliding trajectories in Figs.~\ref{fig:recollision_probability_theta}c and~\ref{fig:recollision_probability_theta}d for $\Theta = 0$ and $\Theta= \pi/2$. 
The blue and magenta curves are RPOs referred to as $\mathcal{O}_2^{\leftarrow}$ and $\mathcal{O}_2^{\uparrow}$, respectively. We observe that, even in 2D, one piece of the RPOs is close to the origin and another piece is far from the origin. If the electron ionizes in the neighborhood of a RPO, like for the 1D case, it is likely that it ionizes in a region of phase space where invariant structures drive the electron and, in particular, drive it back to its parent ion. 
In Figs.~\ref{fig:traj_vs_RPOs}a and~\ref{fig:traj_vs_RPOs}b, we observe that $\mathcal{O}_2^{\uparrow}$ is the closest to the origin when $\omega t+\varphi = 3\pi/2$. At this time, its momentum is along $\mathbf{e}_x$. As a consequence, $\mathcal{O}_2^{\uparrow}$ is used for obtaining the recollision conditions when the XUV angle is $\Theta = 0$ and $\pi$ (XUV angle for which the initial momentum of the electron is aligned with the momentum of $\mathcal{O}_2^{\uparrow}$ when it is the closest from the origin). Also, given the invariance of Hamiltonian~\eqref{eq:main_hamiltonian} under the transformation~\eqref{eq:symmetry1}, there exists a RPO $\mathcal{O}_2^{\downarrow}$ which is symmetric to $\mathcal{O}_2^{\uparrow}$ and which is the closest to the origin at $\omega t+\varphi = \pi/2$. In Figs.~\ref{fig:traj_vs_RPOs}c and~\ref{fig:traj_vs_RPOs}d, we observe that $\mathcal{O}_2^{\leftarrow}$ is the closest to the origin when $\omega t+\varphi = 0$. At this time, its momentum is along $\mathbf{e}_y$. As a consequence, $\mathcal{O}_2^{\leftarrow}$ is used for obtaining the recollision conditions when the XUV angle is $\Theta = \pi/2$ and $3\pi/2$ (XUV angle for which the initial momentum of the electron is aligned with the momentum of $\mathcal{O}_2^{\leftarrow}$ when it is the closest from the origin). Also, given the invariance of Hamiltonian~\eqref{eq:main_hamiltonian} under the transformation~\eqref{eq:symmetry1}, there exists a RPO $\mathcal{O}_2^{\rightarrow}$ which is symmetric to $\mathcal{O}_2^{\leftarrow}$ and which is the closest to the origin at $\omega t+\varphi = \pi$. 
In Fig.~\ref{fig:recollision_probability_theta}, the blue asterisks and magenta crosses are the energy of the RPOs given by Hamiltonian~\eqref{eq:main_hamiltonian} when it is the closest from the origin. The abscissa of the markers indicate for which XUV angles the initial momentum of the electron~\eqref{eq:initial_momentum_XUV} is aligned with the momentum of the RPOs when they are the closest from the origin. This condition is necessary for the electron to be close to the RPO after ionization. We observe that the location of the RPOs in phase space indicates well the initial energies of the electrons for which it is the most likely to return to its parent ion. 

\begin{figure}
	\centering
	\includegraphics[width=\textwidth]{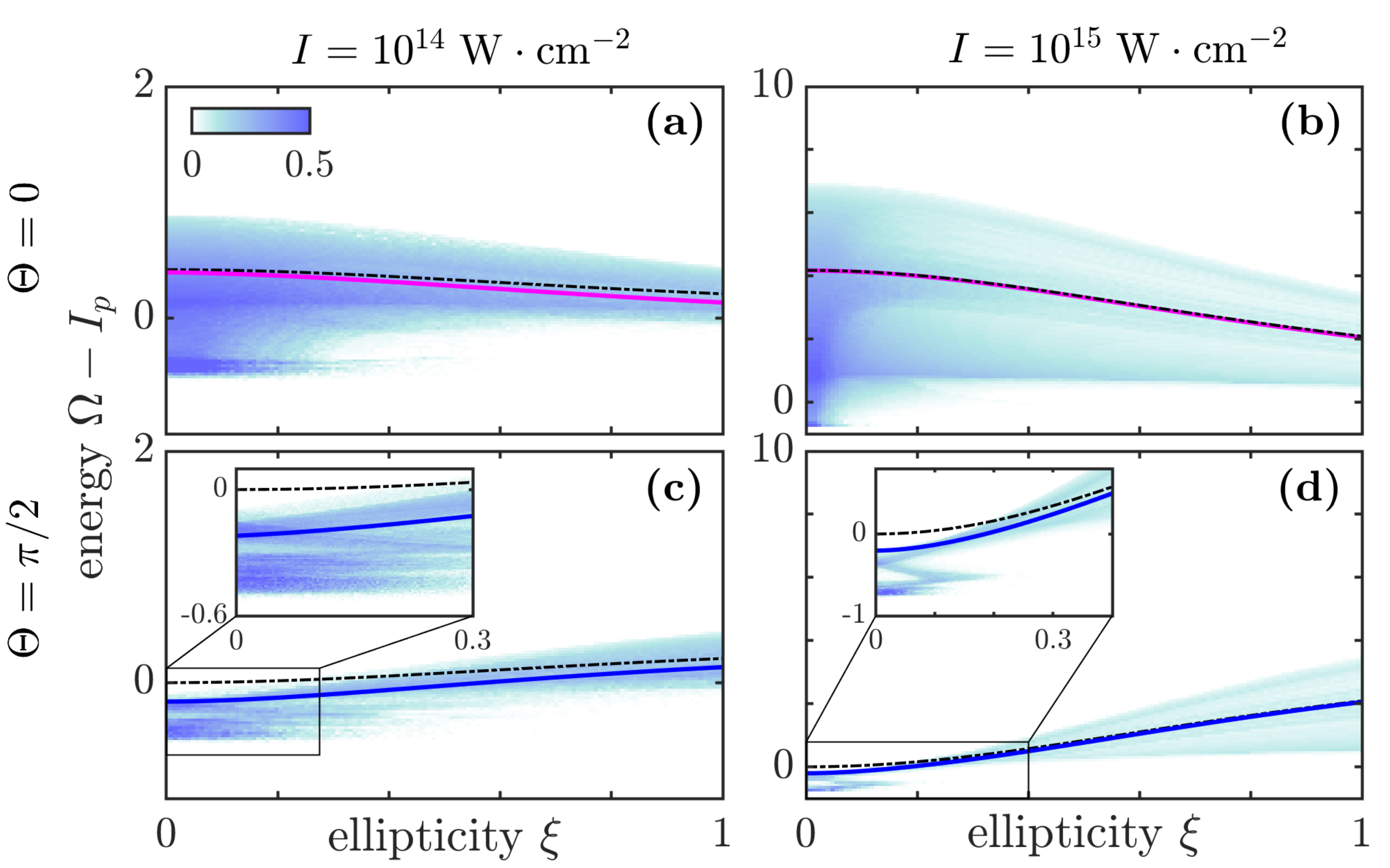}
	\caption{Recollision probabilities of Hamiltonian~\eqref{eq:main_hamiltonian} averaged over the CEP $\varphi$ as a function of the initial energy of the electron $E = \Omega - I_p$ and the ellipticity of the IR field $\xi$ for an integration time of $24 T$. The dash-dotted black curve corresponds to the initial energies and XUV angles which fulfill the recollision conditions in the SFA~\eqref{eq:recollision_condition_SFA}. The magenta and blue curves correspond to the energy of the RPOs $\mathcal{O}_2^{\uparrow}$ and $\mathcal{O}_2^{\rightarrow}$, respectively, at the closest point from the origin as a function of the ellipticity of the IR field. The insets are zoom in the regions indicated by black rectangles. Energies are in a.u.}
	\label{fig:recollision_probability}
\end{figure}

Finally, Fig.~\ref{fig:recollision_probability} shows the recollision probabilities of Hamiltonian~\eqref{eq:main_hamiltonian} as a function of the XUV frequency $\Omega$ and the ellipticity of the IR field $\xi$. The initial position of the electrons is $\mathbf{r}_0 = \boldsymbol{0}$ and their initial momentum is given by Eq.~\eqref{eq:initial_momentum_XUV} for $\Theta = 0$ and $\Theta = \pi/2$. The magenta and blue curves correspond to the energy of the RPOs $\mathcal{O}_2^{\uparrow}$ and $\mathcal{O}_2^{\rightarrow}$ which co-rotate with the IR field. The continuation method employed to follow the RPOs as a function of the parameters of the IR field is described in Appendix~\ref{app:continuation_method}. First, we find that the RPOs $\mathcal{O}_2^{\uparrow}$ and $\mathcal{O}_2^{\rightarrow}$ which co-rotate with the IR field exist regardless of the ellipticity of the IR field. The persistence of these RPOs as a function of the ellipticity of the IR field is studied in Sec.~\ref{sec:elliptically_polarized}.
In Fig.~\ref{fig:recollision_probability}, we observe that the initial energies for which the probability the electron undergoes recollisions is large follows the blue and magenta curves. Hence, the conditions under which the electron undergoes recollisions is well described by the location of RPOs in phase space, regardless of the ellipticity of the IR field. In particular, in the inset of Fig.~\ref{fig:recollision_probability}c and~\ref{fig:recollision_probability}d,  for low ellipticities (i.e., for $\xi \lesssim 0.3$), the blue and magenta curves predict well these conditions while the recollision conditions in the SFA~\eqref{eq:recollision_condition_SFA} cannot predict them accurately, in particular because the energy of the electron in the SFA is always positive.

\section{Primary and secondary RPOs \label{sec:Origin_RPOs}}

In this section, we determine the provenance of the RPOs. First, we consider the field-free atom ($I=0$). In the 1D case ($d=1$), we identify the primary RPO $\mathcal{O}_1$ from the family $\mathcal{O}_1$. In the 2D case ($d=2$), we identify the primary RPO $\mathcal{O}_2$ from the family $\mathcal{O}_2$, which is outside the 1D invariant subspace. We note that the RPOs are labeled by $\mathcal{O}_1$ if it comes from the 1D case in the field-free atom and $\mathcal{O}_2$ if it comes from the 2D case in the field-free atom. Second, we consider $\xi = 0$ and we study the linear stability of $\mathcal{O}_1$ and $\mathcal{O}_2$ as a function of the intensity of the IR field. From the bifurcation diagrams, we identify the secondary RPOs of $\mathcal{O}_1$ and $\mathcal{O}_2$, namely $\mathcal{O}_1^{\leftrightarrows}$, $\mathcal{O}_2^{\leftrightarrows}$ and $\mathcal{O}_2^{\uparrow\downarrow}$. The RPOs are labeled with arrows to indicate their location in the plane $(x,y)$ with respect to the origin. For instance, we observe in Fig.~\ref{fig:traj_vs_RPOs}a that the magenta curve is above the origin: This RPO is labeled as $\mathcal{O}_2^{\uparrow}$. The labels of the primary RPOs $\mathcal{O}_1$ and $\mathcal{O}_2$ have no arrows since they are symmetric with respect to the origin (see for instance Fig.~\ref{fig:traj1D_vs_RPOs}, and Figs.~\ref{fig:RPOs_OA_OF_py}a and~\ref{fig:RPOs_OA_OF_py}b). The labels with two arrows refer to the two symmetric RPOs, for instance $\mathcal{O}_2^{\uparrow\downarrow}$ refers to $\mathcal{O}_2^{\uparrow}$ and $\mathcal{O}_2^{\downarrow}$. Second, we show that for very high intensity, the RPO family $\mathcal{O}_2$ becomes part of the RPO family $\mathcal{O}_1$. 

\subsection{Field-free atom: Primary RPOs $\mathcal{O}_1$ and $\mathcal{O}_2$ \label{sec:field-free-atom}}
\subsubsection{1D case ($d=1$): Primary RPO $\mathcal{O}_1$ \label{sec:1D_RPO_O1}}
In the 1D case and in the field-free atom (for $I = 0$), Hamiltonian~\eqref{eq:main_hamiltonian} becomes
\begin{equation*}
H (x,p_x) = \dfrac{p_x^2}{2} - \dfrac{1}{\sqrt{x^2+1}} .
\end{equation*}
The energy of the electron $E = H(x,p_x)$ is conserved. For $E<0$, the trajectories are bounded and periodic. In order to determine periodic orbits in the field-free atom which persist when the laser field is turned on, we identify the ones for which the period is the same as the period of the IR field $T = 2\pi/\omega$. In the field-free atom, there is a unique energy $E$ for which the period of the periodic orbit is $T$. This energy $E$ is solution of the equation
\begin{equation*}
T = 4  \int_0^{x_m (E)} \mathrm{d}x \left[ 2 \left( E + \dfrac{1}{\sqrt{x^2+1}} \right) \right]^{-1/2}  ,
\end{equation*}  
where $x_m (E) = ( 1/E^2 - 1 )^{1/2}$ is the maximum distance between the electron and the origin. For $\omega = 0.0584$, the energy of the periodic orbit of period $T$ is $E \approx - 0.12$ and its maximum distance from the core is $x_m (E) \approx 8.19$ (see Ref.~\cite{Mauger2012_PRE} for an approximation of the energy of the electron in the field-free atom as a function of its period). This periodic orbit corresponds to the RPO $\mathcal{O}_1$. It persists when the laser field is turned on, remains in the invariant subspace for LP fields and is symmetric with respect to the origin. It corresponds to the RPO identified in Ref.~\cite{Kamor2014}, used to describe a 1D recollision scenario.  In Ref.~\cite{Kamor2014}, $\mathcal{O}_1$ has been identified from the SFA. In contrast, here, $\mathcal{O}_1$ has been identified from the field-free atom in 1D ($d=1$). 

\subsubsection{2D case ($d=2$): Primary RPO $\mathcal{O}_2$}
We consider the electron dynamics in 2D, in the polarization plane and in the field-free atom (for $I=0$). We use polar coordinates, where the position of the electron is $\mathbf{r} = \mathbf{e}_x r \cos \nu + \mathbf{e}_y  r \sin \nu$, with $r$ the distance of the electron to the origin and $\nu$ the angle between its position and the $\mathbf{e}_x$-axis. The momenta canonically conjugate to $r$ and $\nu$ are the radial momentum $p_r$ and the angular momentum $p_{\nu}$, respectively. The angular momentum is conserved due to the rotational invariance of the ion-electron potential, i.e., $p_{\nu} = \ell$. The dynamics in the field-free atom is described by the reduced Hamiltonian
\begin{equation*}
H (r,p_r) = \dfrac{p_r^2}{2} + \dfrac{\ell^2}{2 r^2} - \dfrac{1}{\sqrt{r^2+1}} .
\end{equation*}
The energy of the electron is also conserved, i.e., $E = H(r,p_r)$. For $E<0$, the trajectories are bounded. However, in contrast to the hard-Coulomb potential (for which $V (\mathbf{r}) = -1/|\mathbf{r}|$), the radial frequency is different than the angular frequency in action-angle variables~\cite{Goldstein}. As a consequence, the orbits are not necessarily periodic. We consider the circular periodic orbits, for which $\dot{r} = p_r = 0$ and $\dot{\nu} = \ell / R^2$ is conserved, where $R$ is the radius of the circular periodic orbit. Like in the 1D case (see Sec.~\ref{sec:1D_RPO_O1}) we consider the periodic orbits of period $T$. Therefore the frequency is $\dot{\nu} = \pm \omega$ and the angular momentum is $\ell = \pm R^2 \omega$. The radius of the circular periodic orbits $R$ is solution of the equation $\dot{p}_r = 0$, leading to $\ell^2 = R^4 (R^2+1)^{-3/2}$. As a consequence, the radius of the circular orbits of period $T$ is $R^2 = \omega^{-4/3} - 1$,  their angular momentum is $\ell = \pm R^2 \omega$ and their energy is
\begin{equation*}
E = - \dfrac{\omega^{2/3}}{2} \left( 1 + \omega^{4/3} \right)  .
\end{equation*}
For $\omega = 0.0584$, the energy of the circular periodic orbit of period $T$ is $E \approx - 0.078$, its radius is $R \approx 8.35$ and its angular momentum is $\ell \approx \pm 4.07$. These periodic orbits correspond to the RPOs $\mathcal{O}_2$. They originate from the field-free atom in 2D ($d=2$). They persist when the laser field is turned on and are symmetric with respect to the origin. Here, we observe that there are two RPOs $\mathcal{O}_2$, one with positive angular momentum (i.e., $\ell >0$) and one with negative angular momentum (i.e., $\ell <0$). 
As a consequence, as mentioned in Sec.~\ref{sec:RPOs_return_conditions}, when the IR field is turned on and $\xi \neq 0$, there is one periodic orbit which co-rotates with the IR field and one which counter-rotates with the IR field. We show in Sec.~\ref{sec:elliptically_polarized} that only $\mathcal{O}_2$ which co-rotates with the IR field persists for large ellipticities and large intensities of the IR field, while the one which counter-rotates with the IR field collapses with $\mathcal{O}_1$ at a given ellipticity for strong intensity fields.
We note that $\mathcal{O}_2$ is relatively far from the core (more than around ten atomic units). Given that the initial position of the electron is $\mathbf{r}_0 = \boldsymbol{0}$, it ionizes in a region of phase space far from the one where $\mathcal{O}_2$ is located. Therefore $\mathcal{O}_2$ does not appear in Figs.~\ref{fig:recollision_probability_theta} and~\ref{fig:recollision_probability}. In this study, the main importance of this RPO is that as the intensity of the IR field increases, $\mathcal{O}_2$ undergoes bifurcations and leads to RPOs which are close to the origin (in particular $\mathcal{O}_2^{\leftrightarrows}$ and $\mathcal{O}_2^{\uparrow\downarrow}$).

\subsection{Bifurcation diagrams for LP IR fields: Secondary RPOs $\mathcal{O}_1^{\leftrightarrows}$, $\mathcal{O}_2^{\leftrightarrows}$ and $\mathcal{O}_2^{\uparrow\downarrow}$ \label{sec:LP_fields}}

\begin{figure}
	\centering
	\includegraphics[width=\textwidth]{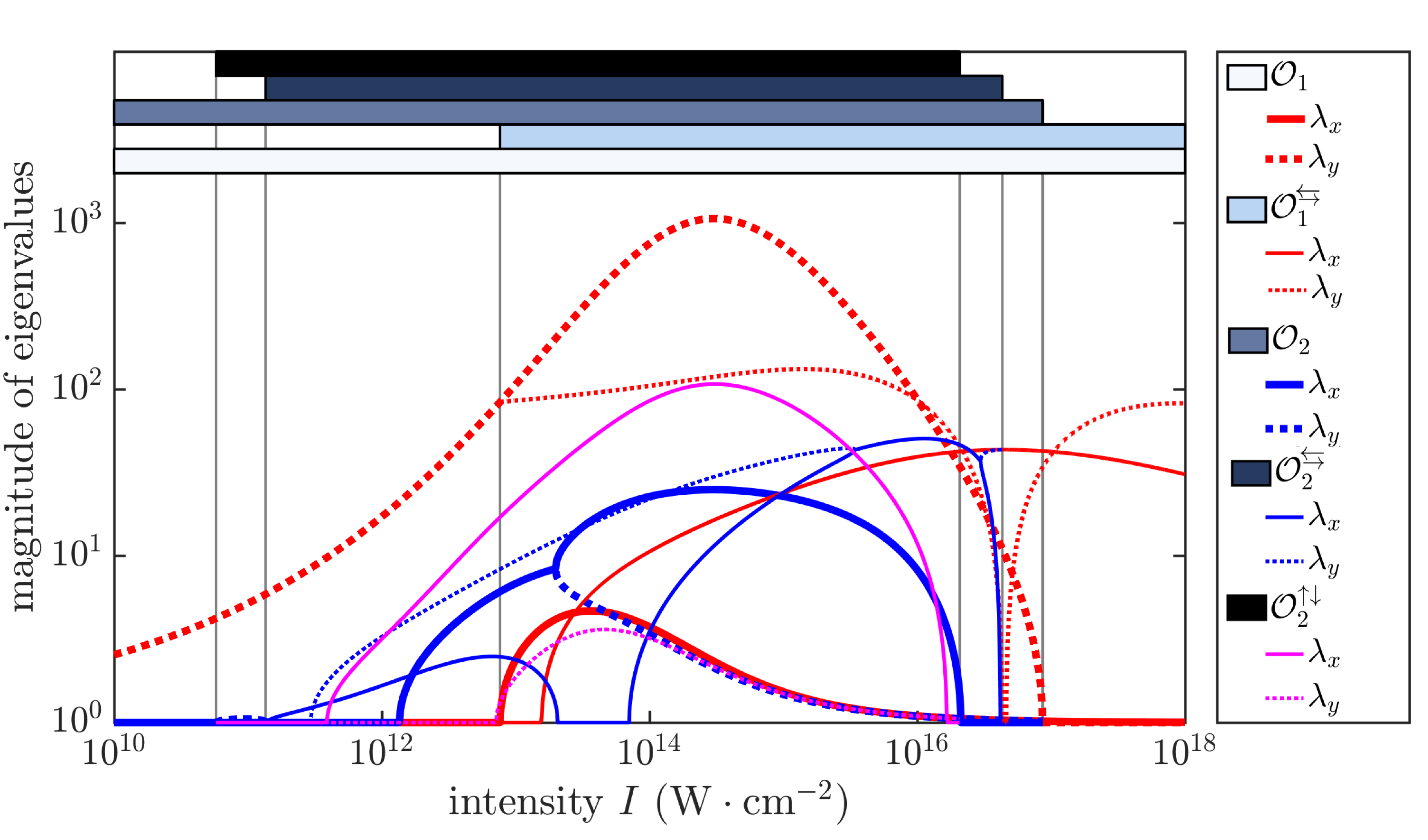}
	\caption{Magnitude of the two largest eigenvalues (in modulus) of the RPOs of Hamiltonian~\eqref{eq:main_hamiltonian} as a function the laser intensity $I$ for 2D ($d = 2$). Because of symmetries, the eigenvalues of $\mathcal{O}_1^{\rightarrow}$ and $\mathcal{O}_1^{\leftarrow}$ are the same, the eigenvalues of $\mathcal{O}_2^{\rightarrow}$ and $\mathcal{O}_2^{\leftarrow}$ are the same, and the eigenvalues of $\mathcal{O}_2^{\uparrow}$ and $\mathcal{O}_2^{\downarrow}$ are the same. The filled rectangles and the vertical solid gray lines indicate the range of intensity of the IR field for which the RPOs exist. For $\mathcal{O}_1$ and $\mathcal{O}_1^{\leftrightarrows}$, the solid (resp. dotted) lines indicate the magnitude of the eigenvalues $|\lambda_x|$ (resp. $|\lambda_y|$) associated with the eigenvectors in (resp. transverse to) the invariant subspace $(x,p_x)$. The magnitude of the eigenvalues is in a.u.}
	\label{fig:eigenvalues}
\end{figure}

We consider LP IR fields ($\xi = 0$) and $d=2$. We use a continuation method described in Appendix~\ref{app:continuation_method} to follow the periodic orbits $\mathcal{O}_1$ and $\mathcal{O}_2$, identified in Sec.~\ref{sec:field-free-atom}, as a function of the intensity of the IR field $I$. Figure~\ref{fig:eigenvalues} shows the magnitude of the two largest eigenvalues (in modulus) of the monodromy matrix of the RPOs of the family $\mathcal{O}_1$ (red thick curves) and $\mathcal{O}_2$ (blue thick curves) as a function of the intensity of the IR field. For the family $\mathcal{O}_1$, the solid (resp. dotted) curves are the magnitude of the eigenvalue associated with the eigenvector in (resp. transverse to) the invariant subspace $(x,p_x)$. 
The largest eigenvalue of the RPOs provides the characteristic time during which an electron initiated in the neighborhood of the RPO stays close to it. For $|\lambda| = 1$ (resp. $\lambda > 1$) the electron follows a stable orbit (resp. unstable orbit) and therefore stays close to (resp. goes away from) the neighborhood of the RPO. 
In Fig.~\ref{fig:eigenvalues}, for intensities $I \in [10^{12} , 10^{16}] \; \mathrm{W} \ \mathrm{cm}^{-2}$, we observe that the largest eigenvalue of $\mathcal{O}_1$ is about $10$ times larger than the largest eigenvalue of $\mathcal{O}_2$. The largest eigenvalue of $\mathcal{O}_1$ reaches $|\lambda_y| \sim 10^{3}$ and its associated eigenvector is in the direction transverse to the invariant subspace $(x,p_x)$. As a consequence, in this range of intensity of the IR field, the trajectories initiated near $\mathcal{O}_1$ and outside the invariant subspace $(x,p_x)$ are pushed far away from it rather quickly (around $10^{3}$ a.u. per unit of time). Despite the existence of this invariant subspace for LP fields, two-dimensional RPOs also play a role for the determination of the return conditions of the electron.
\par
For increasing intensities of the IR field, we observe that the RPOs $\mathcal{O}_1$ (red thick curves in Fig.~\ref{fig:eigenvalues}) and $\mathcal{O}_2$ (blue thick curves in Fig.~\ref{fig:eigenvalues}) undergo multiple bifurcations. In particular, for $I \sim 10^{13} \; \mathrm{W} \ \mathrm{cm}^{-2}$, $\mathcal{O}_1$ undergoes a \emph{pitchfork bifurcation} in the invariant subspace $(x,p_x)$ due to the symmetry~\eqref{eq:symmetry1}. At this intensity, the secondary RPOs $\mathcal{O}_1^{\leftrightarrows}$ (red thin curves in Fig.~\ref{fig:eigenvalues}) appear. They are shown in phase space in Fig.~\ref{fig:traj1D_vs_RPOs}. After the bifurcation, $\mathcal{O}_1^{\leftrightarrows}$ are stable in the invariant subspace $(x,p_x)$. The RPO $\mathcal{O}_1^{\leftarrow}$ is symmetric to $\mathcal{O}_1^{\rightarrow}$ with respect to the origin according to the invariance over the transformation~\eqref{eq:symmetry1}. The RPOs $\mathcal{O}_1^{\leftrightarrows}$ have been used in Ref.~\cite{Norman2015} to describe ionization stability, in Ref.~\cite{Berman2015} to asses the persistence of Coulomb focusing for very intense laser fields, and in Ref.~\cite{Abanador2017} to describe HHG with elliptically polarized fields. 
Similarly, for $I \sim 10^{11} \; \mathrm{W} \ \mathrm{cm}^{-2}$, the RPO $\mathcal{O}_2$ undergoes two successive bifurcations which give rise to $\mathcal{O}_2^{\uparrow\downarrow}$ (magenta thin curves in Fig.~\ref{fig:eigenvalues}) and $\mathcal{O}_2^{\leftrightarrows}$ (blue thin curves in Fig.~\ref{fig:eigenvalues}). The RPOs $\mathcal{O}_2^{\uparrow}$ and $\mathcal{O}_2^{\leftarrow}$ are symmetric to $\mathcal{O}_2^{\downarrow}$ and $\mathcal{O}_2^{\rightarrow}$, respectively, with respect to the origin according to the invariance under the rotation~\eqref{eq:symmetry1}. 
The larger the intensity and the closer along the $\mathbf{e}_x$-axis get the RPOs of the family $\mathcal{O}_2$. For very high intensities, at $I \sim [10^{16} , 10^{17}] \; \mathrm{W} \ \mathrm{cm}^{-2}$, the RPOs of the family $\mathcal{O}_2$ collapse in the invariant subspace $(x,p_x)$ and disappear, i.e., they have no component along the $\mathbf{e}_y$-axis and become part of the family $\mathcal{O}_1$.
In Fig.~\ref{fig:eigenvalues}, we observe that the red thick dotted curve undergoes a bifurcation when $\mathcal{O}_2$ becomes $\mathcal{O}_1$, the red thick solid curve undergoes a bifurcation when $\mathcal{O}_2^{\uparrow\downarrow}$ becomes $\mathcal{O}_1$, and the red thin dotted curve undergoes a bifurcation when $\mathcal{O}_2^{\leftrightarrows}$ becomes $\mathcal{O}_1^{\leftrightarrows}$.
\par
Finally, due to the symmetry~\eqref{eq:symmetry3} for LP fields, there are two distinct RPOs for each $\mathcal{O}_2$ and $\mathcal{O}_2^{\leftrightarrows}$. For LP fields, when represented in the plane $(x,y)$, $\mathcal{O}_2$ or $\mathcal{O}_2^{\leftrightarrows}$ can be traveled clockwise or anticlockwise. This degeneracy is removed if $\xi > 0$. For the rest of this article, we refer to as $\mathcal{O}_{2 \rm c}^{\cdot}$ and $\mathcal{O}_{2 \rm a}^{\cdot}$ the RPOs of the family $\mathcal{O}_{2}$ which co-rotates and counter-rotates with the EP field~\eqref{eq:laser_field}, respectively.

\section{Persistence of RPOs for increasing ellipticity of the IR field \label{sec:elliptically_polarized}}

\begin{figure}
	\centering
	\includegraphics[width=\textwidth]{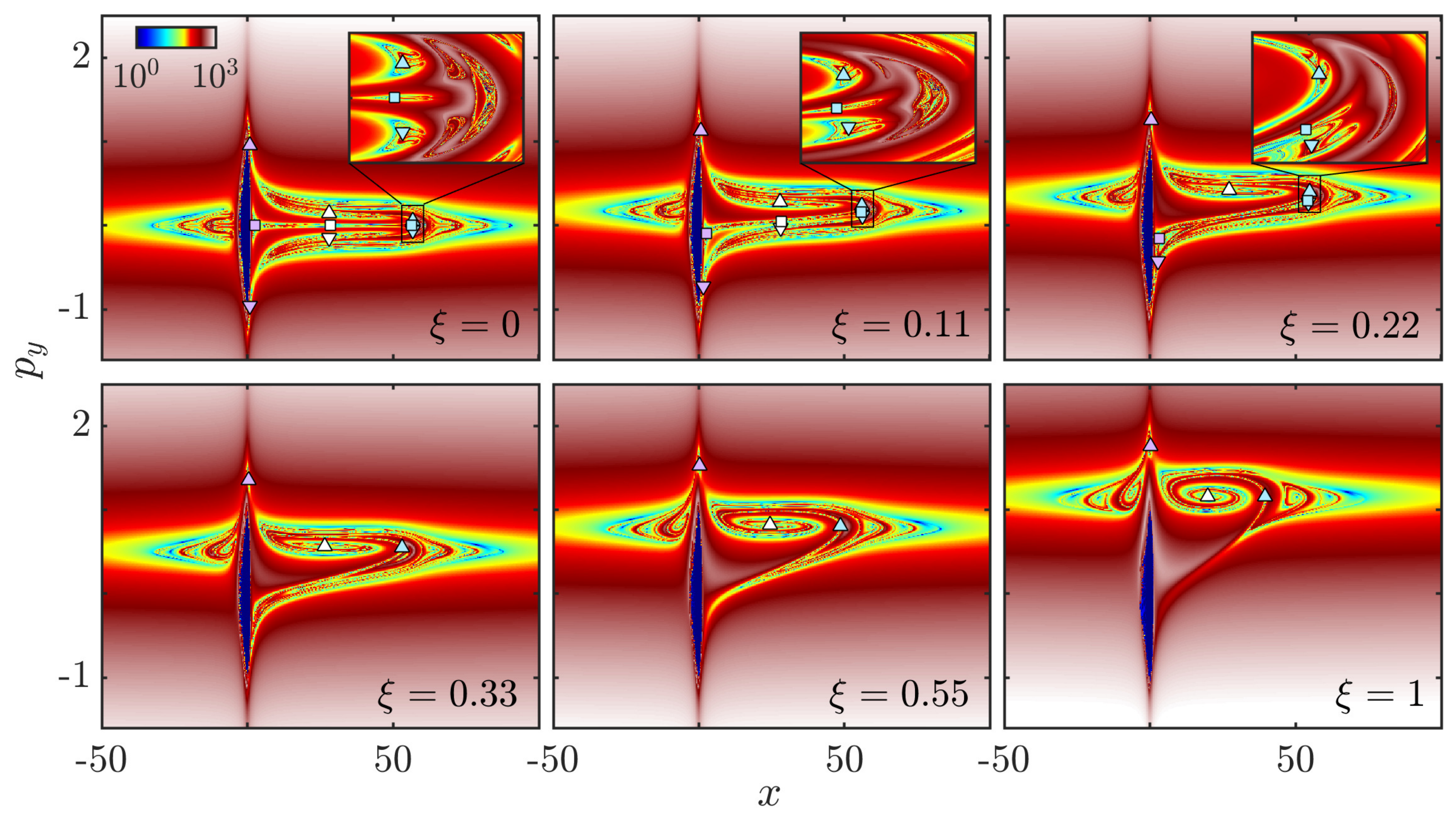}
	\caption{Final distance of the electron of Hamiltonian~\eqref{eq:main_hamiltonian} as a function of the initial conditions in the plane $(x,p_y)$ at time $\omega t + \varphi =0$ for $I = 3 \times 10^{14} \; \mathrm{W} \cdot \mathrm{cm}^{-2}$ for an integration time of $10T$. The marks indicate the RPOs at time $\omega t + \varphi = 0$: Squares are the family $\mathcal{O}_1$, up-pointing triangles are the family $\mathcal{O}_{2}$ co-rotating with the IR field and down-pointing triangles are the family $\mathcal{O}_{2}$ counter-rotating with the IR field. From left to right, it corresponds to the secondary orbits $\mathcal{O}_1^{\leftarrow}$ and $\mathcal{O}_2^{\leftarrow}$ (in light purple), the primary orbits $\mathcal{O}_1$ and $\mathcal{O}_2$ (in white) and the secondary orbits $\mathcal{O}_1^{\rightarrow}$ and $\mathcal{O}_2^{\rightarrow}$ (in light blue). Positions and momenta are in a.u.}
	\label{fig:OA_OF_xpy_ellipticity}
\end{figure}

Figure~\ref{fig:OA_OF_xpy_ellipticity} shows the final distance of the electron as a function of its initial conditions $(x,p_y)$ for $y = p_x = 0$ for an integration time of 10 laser cycles and $I = 3\times 10^{14} \; \mathrm{W}\cdot\mathrm{cm}^{-2}$. We observe sensitivity with respect to the initial conditions as a signature of recollisions~\cite{Dubois2019}. We refer to these regions as \emph{recolliding regions} in phase space. The XUV frequency $\Omega$ and angle $\Theta$ for which the electron undergoes recollisions are the ones for which the electron is initiated in the recolliding regions [where $\mathbf{r}_0 = \boldsymbol{0}$ and $\mathbf{p}_0$ given by Eq.~\eqref{eq:initial_momentum_XUV}].
In Fig.~\ref{fig:OA_OF_xpy_ellipticity}, we observe patterns for the recolliding regions in phase space. In the middle of these recolliding regions in phase space are located the RPOs $\mathcal{O}_1$, $\mathcal{O}_1^{\leftrightarrows}$, $\mathcal{O}_2$ and $\mathcal{O}_2^{\leftrightarrows}$, which belong to the plane $(x,p_y)$ (i.e., $y = p_x = 0$) for $\omega t + \varphi = 0$ and for all ellipticities of the IR field. The location of RPOs in phase space are a robust indication of the location of recolliding regions.
\par
In Fig.~\ref{fig:OA_OF_xpy_ellipticity}, for LP fields ($\xi = 0$), we observe that the recolliding regions are symmetric with respect to the $\mathbf{e}_y$-axis due to the invariance under the transformation~\eqref{eq:symmetry3}.
For small ellipticities (i.e., $\xi \lesssim 0.5$), we observe mainly two branches in this region, one branch where $p_y$ is negative and one branch where $p_y$ is positive. We observe that $\mathcal{O}_{1}$ and $\mathcal{O}_{1}^{\leftrightarrows}$, and the counter-rotating RPOs $\mathcal{O}_{2 \rm a}$ and $\mathcal{O}_{2 \rm a}^{\leftrightarrows}$ are in the branch where $p_y$ is negative. 
The co-rotating RPOs $\mathcal{O}_{2 \rm c}$ and $\mathcal{O}_{2 \rm c}^{\leftrightarrows}$ are in the branch where $p_y$ is positive. For increasing ellipticities, the recolliding regions move to locations in phase space where the momentum $p_y$ is larger. The area of the branch where $p_y$ is positive is roughly the same, while the area of the branch where $p_y$ is negative decreases. The area of the latter branch becomes almost negligible at $\xi = 0.33$, when the RPOs $\mathcal{O}_1$ and counter-rotating RPOs $\mathcal{O}_{2 \rm a}$ disappear. This shows that, as mentioned in the previously, the location of the RPOs in phase space indicates the location of recolliding regions.
\par
In this section, we determine the mechanism behind the disappearance of the RPOs $\mathcal{O}_1$ and $\mathcal{O}_{2 \rm a}$ as ellipticity increases. We show that, for the range of intensities studied in this article, only the RPOs $\mathcal{O}_{2 \rm c}$ exist regardless of the laser ellipticity for strong intensity of the IR field. Then, we show the relation between the counter-rotating RPOs $\mathcal{O}_{2 \rm c}$ and $\mathcal{O}_{2 \rm c}^{\leftrightarrows}$ with the ones identified in Refs.~\cite{Kamor2013, Mauger2014_JPB} for CP fields in the rotating frame (RF).

\subsection{Bifurcation diagrams for EP IR fields}
 
\begin{figure}
	\centering
	\includegraphics[width=\textwidth]{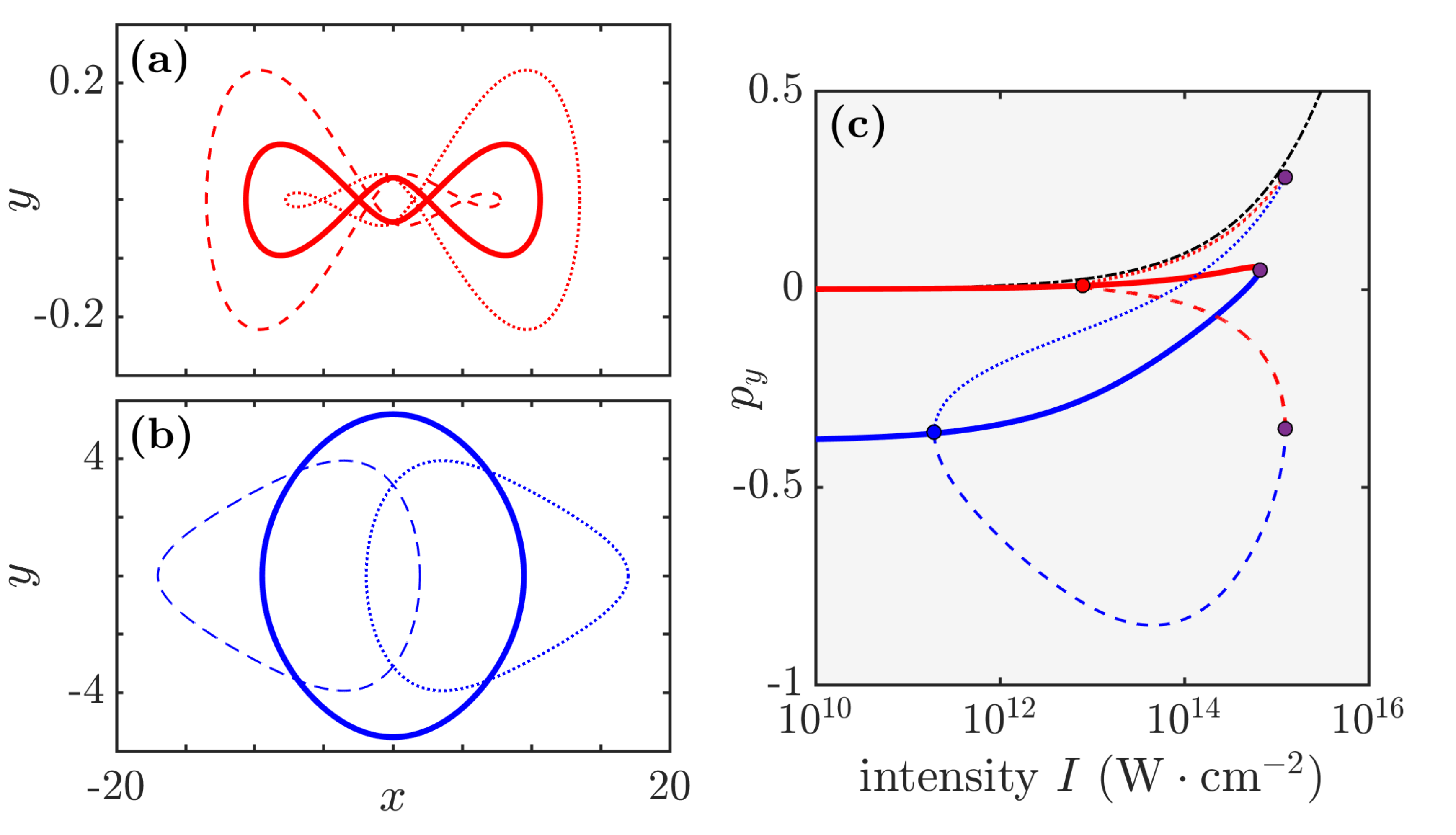}
	\caption{RPOs of Hamiltonian~\eqref{eq:main_hamiltonian} and their transverse momentum $p_y$ as a function of the laser intensity for $\xi = 0.1$. (a) Periodic orbits $\mathcal{O}_1$ (solid lines), $\mathcal{O}_1^{\rightarrow}$ (dotted lines) and $\mathcal{O}_1^{\leftarrow}$ (dashed lines). (b) Periodic orbits $\mathcal{O}_{2 \rm a}$ (solid lines), $\mathcal{O}_{2 \rm a}^{\rightarrow}$ (dotted lines) and $\mathcal{O}_{2 \rm a}^{\leftarrow}$ (dashed lines) counter-rotating with the laser field. In panels (a) and (b), the laser intensity is $I = 10^{13} \; \mathrm{W} \cdot \mathrm{cm}^{-2}$. (c) Momentum $p_y$ of the fixed point of the periodic orbits for $\omega t + \varphi = 0$ (which corresponds to the time at which the $x$-component of an electron trajectory along the RPO reaches the maximum) as a function of the laser intensity. The color code is the same as in the panels (a) and (b). The red, blue and purple dots show the turning points of the continuation method. The black dash-dotted curve indicates $p_y = \xi E_0/\omega\sqrt{\xi^2+1}$. Positions and momenta are in a.u.}
	\label{fig:RPOs_OA_OF_py}
\end{figure}

In order to determine the mechanism behind the disappearance of the RPOs $\mathcal{O}_1$, $\mathcal{O}_1^{\leftrightarrows}$, $\mathcal{O}_{2 \rm a}$, $\mathcal{O}_{2 \rm a}^{\leftrightarrows}$ for varying laser parameters, we first follow them as a function the intensity of the IR field for $\xi = 0.1$ (see Appendix~\ref{app:continuation_method} for details on the continuation method).
Figure~\ref{fig:RPOs_OA_OF_py}c shows the momentum $p_y$ of these RPOs as a function of the intensity of the IR field for $\omega t + \varphi = 0$ and $\xi = 0.1$. It is comparable to Fig.~\ref{fig:RPOs_OA_OF_py} for LP fields ($\xi = 0$). 
Like for LP fields, for weak intensities (such as when the IR field is turned off), only the primary RPOs $\mathcal{O}_1$ (solid red curves in Figs.~\ref{fig:RPOs_OA_OF_py}a and~\ref{fig:RPOs_OA_OF_py}c) and $\mathcal{O}_{2 \rm a}$ (solid blue curves in Figs.~\ref{fig:RPOs_OA_OF_py}b and~\ref{fig:RPOs_OA_OF_py}c) are present. As the intensity of the IR field increases, the primary RPO $\mathcal{O}_1$ undergoes a Pitchfork bifurcation at $I \sim 10^{13} \; \mathrm{W}\ \mathrm{cm}^{-2}$ (red dot in Fig.~\ref{fig:RPOs_OA_OF_py}c) and $\mathcal{O}_1^{\leftrightarrows}$ (dashed and dotted red curves in Figs.~\ref{fig:RPOs_OA_OF_py}a and~\ref{fig:RPOs_OA_OF_py}c) appear. Similarly, the primary RPO $\mathcal{O}_{2 \rm a}$ undergoes a Pitchfork bifurcation at $I \sim 10^{12} \; \mathrm{W}\ \mathrm{cm}^{-2}$ (blue dot in Fig.~\ref{fig:RPOs_OA_OF_py}c) and $\mathcal{O}_{2\rm a}^{\leftrightarrows}$ (dashed and dotted blue curves in Figs.~\ref{fig:RPOs_OA_OF_py}b and~\ref{fig:RPOs_OA_OF_py}c) appear.
However, in contrast to LP fields, at $I \sim 10^{14} \; \mathrm{W}\ \mathrm{cm}^{-2}$ (purple dots in Fig.~\ref{fig:RPOs_OA_OF_py}c), the primary orbits $\mathcal{O}_1$ and $\mathcal{O}_{2 \rm a}^{\leftrightarrows}$ meet. In the same way, the RPOs $\mathcal{O}_{2 \rm a}^{\leftarrow}$ and $\mathcal{O}_{2 \rm a}^{\rightarrow}$ meet with $\mathcal{O}_{1}^{\leftarrow}$ and $\mathcal{O}_{1}^{\rightarrow}$, respectively. For intensities larger than $I \sim 10^{14} \; \mathrm{W}\ \mathrm{cm}^{-2}$, the RPOs $\mathcal{O}_1$, $\mathcal{O}_{2\rm a}$, $\mathcal{O}_1^{\leftrightarrows}$ and $\mathcal{O}_{2\rm a}^{\leftrightarrows}$ are not present.
\par
In Fig.~\ref{fig:RPOs_OA_OF_py}c, we observe that $\mathcal{O}_1^{\leftarrow}$, $\mathcal{O}_{2 \rm a}^{\leftarrow}$, $\mathcal{O}_1^{\rightarrow}$, $\mathcal{O}_{2 \rm a}^{\rightarrow}$ form a closed loop in the space of the parameters of the laser. For instance, we start the continuation method with an initial guess for $\mathcal{O}_{2 \rm a}^{\rightarrow}$ at $I \sim 10^{12} \; \mathrm{W} \ \mathrm{cm}^{-2}$ (at the right of the blue dot) corresponding to the blue dotted curve in Fig.~\ref{fig:RPOs_OA_OF_py}c. For increasing intensity, its momentum $p_y$ increases until the intensity reaches $I \sim 10^{14} \; \mathrm{W} \ \mathrm{cm}^{-2}$ (upper purple dot). The continuation method goes on the branch of $\mathcal{O}_1^{\rightarrow}$, corresponding to the red dotted curve in Fig.~\ref{fig:RPOs_OA_OF_py}c. Then, the intensity decreases and the momentum $p_y$ of $\mathcal{O}_1^{\rightarrow}$ decreases, until the intensity reaches $I \sim 10^{12} \; \mathrm{W}\ \mathrm{cm}^{-2}$ (red dot). The continuation method goes on the branch of $\mathcal{O}_1^{\leftarrow}$, corresponding to the red dashed curve in Fig.~\ref{fig:RPOs_OA_OF_py}c. Then, the intensity increases and the momentum $p_y$ of $\mathcal{O}_1^{\leftarrow}$ decreases, until the intensity reaches $I \sim 10^{14} \; \mathrm{W}\ \mathrm{cm}^{-2}$ (lower purple dot). The continuation method goes on the branch of $\mathcal{O}_{2 \rm a}^{\leftarrow}$, corresponding to the blue dashed curve in Fig.~\ref{fig:RPOs_OA_OF_py}c. Then, the intensity decreases and the momentum $p_y$ of $\mathcal{O}_{2 \rm a}^{\leftarrow}$ varies, until the intensity reaches $I \sim 10^{12} \; \mathrm{W}\ \mathrm{cm}^{-2}$ (blue dot). The continuation method goes on the branch of $\mathcal{O}_{2 \rm a}^{\rightarrow}$ and it stops after the loop is closed. This loop in phase and parameter space shows the importance of using a specific continuation method in this case (described in Appendix~\ref{app:continuation_method}) which let the intensity of the IR field free.
\par
Despite the primary and secondary RPOs of the family $\mathcal{O}_1$ and $\mathcal{O}_2$ come from different conditions in the field-free atom, they are connected to each others in the space of the parameters of the IR field.

\begin{figure}
	\centering
	\includegraphics[width=\textwidth]{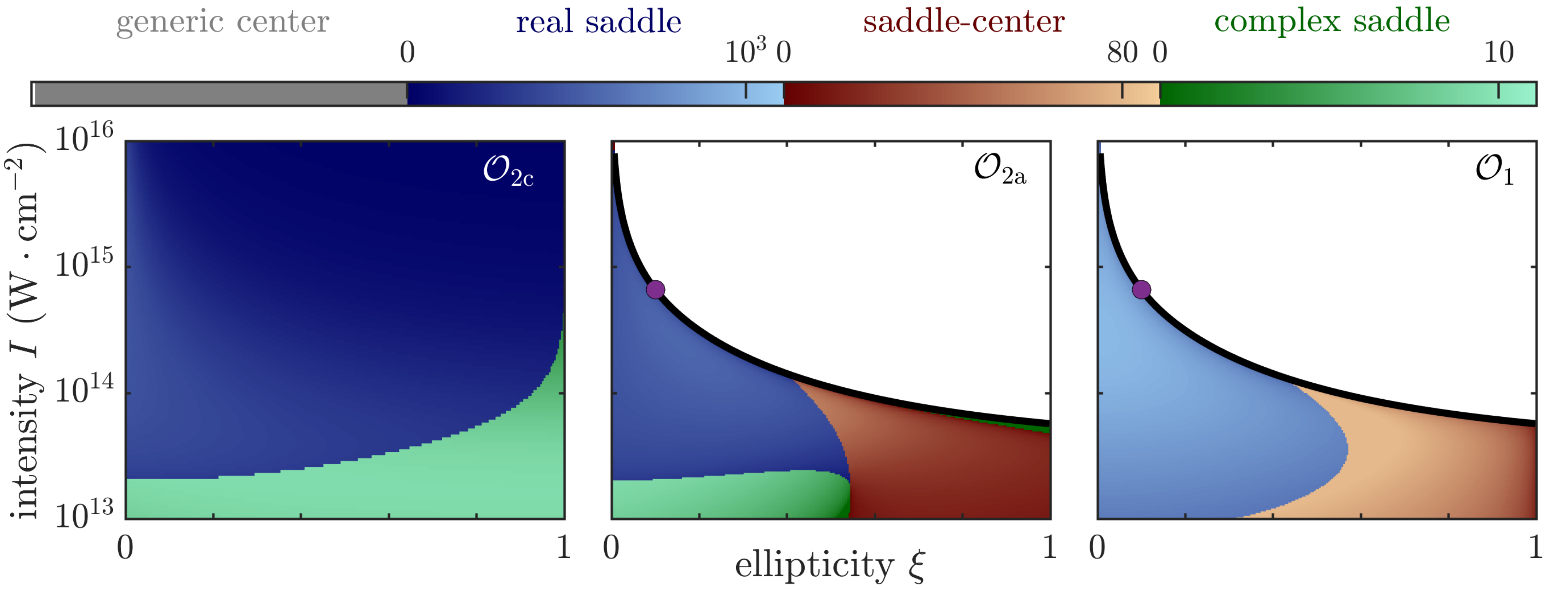}
	\caption{Stability maps of the RPOs as a function of the laser ellipticity $\xi$ and the laser intensity $I$. The color code corresponds to the magnitude of the largest eigenvalue (in modulus) of the RPOs. The black curves indicate the critical intensities (the black curve in the middle panel is the same as the black curve in the right panel). The purple dots correspond to the middle purple dot in Fig.~\ref{fig:RPOs_OA_OF_py}c. The magnitude of the eigenvalues is in a.u.}
	\label{fig:stability_maps}
\end{figure}

Second, we follow these RPOs as a function of the ellipticity and the intensity of the IR field.
Figure~\ref{fig:stability_maps} shows the magnitude of the largest eigenvalue (in modulus) of the monodromy matrix of the RPOs $\mathcal{O}_{2 \rm c}$, $\mathcal{O}_{2\rm a}$ and $\mathcal{O}_1$ as a function of the ellipticity and the intensity of the IR field. Their linear stability \emph{generic center}, \emph{real saddle}, \emph{saddle-center} and \emph{complex saddle} are determined from the analysis of their two largest eigenvalues (in modulus).  First, for $\xi > 0$, we observe that there are critical intensities where $\mathcal{O}_1$ and $\mathcal{O}_{2 \rm a}$ disappear, corresponding to the thick black curves in Fig.~\ref{fig:stability_maps}. These critical intensities are the intensities when $\mathcal{O}_1$ and $\mathcal{O}_{2\rm a}$ meet (also corresponding to the middle purple dot in Fig.~\ref{fig:RPOs_OA_OF_py}c for $\xi = 0.1$).
These RPOs meet when their linear stability are either both real saddle (corresponding to the blue region), or when $\mathcal{O}_1$ is saddle-center (corresponding to the red region) and $\mathcal{O}_{2 \rm a}$ is complex saddle. In both cases, $\mathcal{O}_1$ and $\mathcal{O}_{2 \rm a}$ undergo a \emph{saddle-node bifurcation}. 
\par
In contrast to the RPOs $\mathcal{O}_1$, $\mathcal{O}_{2 \rm a}$, $\mathcal{O}_{2 \rm a}^{\leftrightarrows}$ and $\mathcal{O}_{2 \rm a}^{\uparrow \downarrow}$, the RPOs $\mathcal{O}_{2 \rm c}$, $\mathcal{O}_{2 \rm c}^{\leftrightarrows}$ and $\mathcal{O}_{2 \rm c}^{\uparrow \downarrow}$ persist for a wide range of laser parameters. In particular, $\mathcal{O}_{2 \rm c}$ persists for strong intensities of the IR field (i.e., $I \sim [ 10^{13} , 10^{16} ] \; \mathrm{W}\ \mathrm{cm}^{-2}$) and for all ellipticities. As a consequence, the primary RPO $\mathcal{O}_{2\rm c}$ and the secondary RPOs $\mathcal{O}_{2 \rm c}^{\leftrightarrows}$ and $\mathcal{O}_{2 \rm c}^{\uparrow \downarrow}$ can be used to locate recolliding regions in phase space for varying ellipticities and intensities of the IR field. In particular, in Sec.~\ref{sec:RPOs_return_conditions}, this has allowed us to determine the XUV frequencies and XUV angles for which the electron returns to its parent ion for varying ellipticities of the IR field.

\subsection{CP IR fields and RPOs in the rotating frame \label{sec:RF}}

\begin{figure}
	\includegraphics[width=\textwidth]{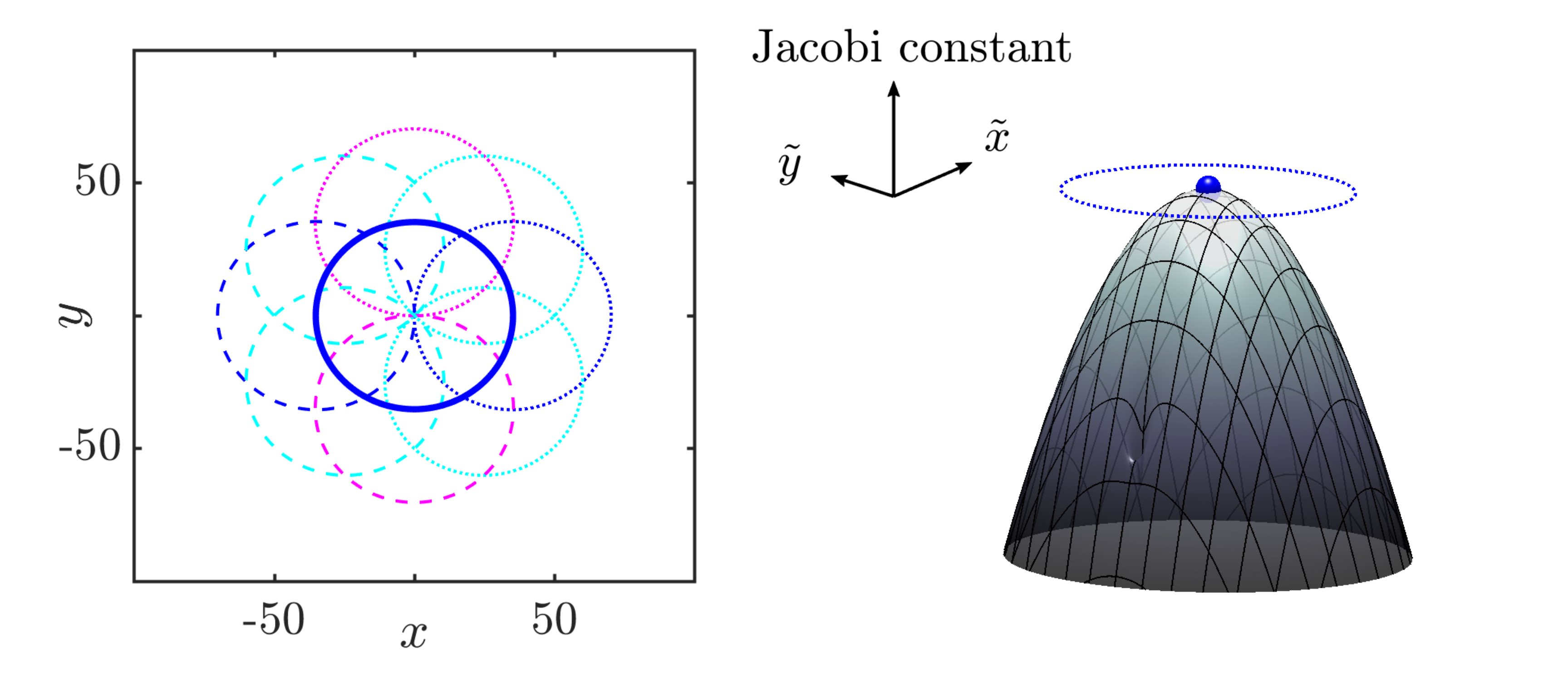}
	\caption{Left panel: RPOs $\mathcal{O}_{2 \rm c}$ (blue solid curve), $\mathcal{O}_{2 \rm c}^{\leftarrow}$ (blue dashed curve), $\mathcal{O}_2^{\rightarrow}$ (blue dotted curve), $\mathcal{O}_{2 \rm c}^{\downarrow}$ (magenta dashed curve) and $\mathcal{O}_2^{\uparrow}$ (magenta dotted curve) for $\xi = 1$ and $I = 10^{15} \; \mathrm{W}\ \mathrm{cm}^{-2}$ in the LF. The cyan dashed and dotted curves are RPOs which are related to $\mathcal{O}_{2 \rm c}^{\leftrightarrows}$ and $\mathcal{O}_{2 \rm c}^{\uparrow\downarrow}$ under the transformation~\eqref{eq:symmetry4}. Right panel: RPOs of the left panel shown in the RF. The blue dotted curve corresponds to $\mathcal{O}_{2 \rm c}^{\leftrightarrows}$, $\mathcal{O}_{2 \rm c}^{\uparrow\downarrow}$ and their symmetric RPOs (they are described by the same orbit in the RF). The gray surface corresponds to the surface where the velocity of the electron vanishes, i.e., $\dot{\tilde{\mathbf{r}}} = \boldsymbol{0}$. The top of the gray surface, indicated by a blue dot, corresponds to $\mathcal{O}_{2 \rm c}$ in the RF. Positions are in a.u.}
	\label{fig:RPOs_LF_RF}
\end{figure}

For CP pulses ($\xi = 1$), due to the invariance of the IR field under the transformation~\eqref{eq:symmetry4}, the energy of the electron in the RF is conserved. The RF corresponds to the framework in which the IR field is static. The energy of the electron in the RF is referred to as the Jacobi constant~\cite{Kamor2013}. The RF is often used to describe the electron dynamics in CP fields, and in particular to describe recollisions~\cite{Mauger2010_PRL,Fu2012,Barrabes2012,Kamor2013,Mauger2014_JPB}. RPOs have been identified in this framework, either in the SFA~\citep{Mauger2010_PRL} or by taking into account the Coulomb interaction~\cite{Kamor2013,Mauger2014_JPB}. The canonical change of coordinates from the laboratory frame (LF) to the RF is given by $\mathbf{r} = \mathbf{R}_z (\omega t) \tilde{\mathbf{r}}$ and $\mathbf{p} = \mathbf{R}_z (\omega t) \tilde{\mathbf{p}}$ where $\mathbf{R}_z (\omega t)$ is the rotation matrix around the $\mathbf{e}_z$-axis of angle $\omega t$, $\tilde{\mathbf{r}}$ is the position of the electron in the RF and $\tilde{\mathbf{p}}$ is its canonically conjugate momentum. 
\par
In the right panel of Fig.~\ref{fig:RPOs_LF_RF}, the gray surface corresponds to the surface where the velocity of the electron vanishes, i.e., $\dot{\tilde{\mathbf{r}}} = \boldsymbol{0}$. The local extrema of the gray surface correspond to fixed points in the RF~\cite{Barrabes2012}. The top of the gray surface, indicated by a blue dot on the right panel of Fig.~\ref{fig:RPOs_LF_RF}, corresponds to the RPO $\mathcal{O}_{2 \rm c}$ in the RF. In the RF, $\mathcal{O}_{2 \rm c}$ is a fixed point, i.e., its phase-space coordinates are such that $\dot{\tilde{\mathbf{r}}} = \dot{\tilde{\mathbf{p}}} = \boldsymbol{0}$, and its distance from the origin is constant. 
In the LF, for CP fields, $\mathcal{O}_{2 \rm c}$ is therefore a circle in the polarization plane (corresponding to the blue thick curve in the left panel of Fig.~\ref{fig:RPOs_LF_RF}), regardless of the laser intensity. For increasing intensities, the radius of $\mathcal{O}_{2 \rm c}$ increases.
\par
In the left panel of Fig.~\ref{fig:RPOs_LF_RF}, the dashed and dotted blue curves are $\mathcal{O}_{2 \rm c}^{\leftrightarrows}$ and the dashed and dotted magenta curves are $\mathcal{O}_{2 \rm c}^{\uparrow\downarrow}$ in the LF for CP fields and $I = 10^{15} \; \mathrm{W}\ \mathrm{cm}^{-2}$.
For CP fields, $\mathcal{O}_{2 \rm c}^{\leftrightarrows}$ is symmetric to $\mathcal{O}_{2 \rm c}^{\uparrow\downarrow}$ under the transformation~\eqref{eq:symmetry4} with $\tau = \pm T/4$. There is an infinity of other RPOs which are symmetric to $\mathcal{O}_{2 \rm c}^{\leftrightarrows}$ and $\mathcal{O}_{2 \rm c}^{\uparrow\downarrow}$ under the transformation~\eqref{eq:symmetry4} for $\tau \in [0,T]$. The cyan dashed and dotted curves correspond to four of these RPOs. In the RF, these RPOs follow the blue dotted curve in the right panel of Fig.~\ref{fig:RPOs_LF_RF}. In the RF, all these RPOs have the same shape and correspond to the same RPO. They correspond to the RPO of period $T$ which goes one loop around the origin identified in Ref.~\cite{Kamor2013}. These RPOs have been used in Refs.~\cite{Kamor2013, Dubois2020} to describe NSDI in CP fields and in Ref.~\cite{Mauger2014_JPB} to describe HHG in CP fields.

\section{Conclusions}

In summary, we have considered the dynamics of an electron in an IR EP field after its ionization by a XUV pulse of frequency $\Omega$ and angle $\Theta$. The identification and the study of RPOs in phase space has allowed us to determine the frequencies and angles of the XUV pulse for which the electron undergoes recollisions. We have shown that regardless of the target atom and the ellipticity of the IR field, there exist frequencies and angles of the XUV for which the electron returns to the parent ion. This suggests that pump-probe experiments are an appropriate setup in order to study the conditions under which the electron in atoms subjected to EP fields undergoes recollisions. 
\par
We have identified two families of RPOs, namely $\mathcal{O}_1$ and $\mathcal{O}_2$. The primary RPOs of these families, namely $\mathcal{O}_1$ and $\mathcal{O}_2$, have been identified from the field-free atom (see Sec.~\ref{sec:1D_RPO_O1}). For increasing intensities of the IR field, we have shown that these two primary RPOs undergo bifurcations, and give rise to secondary RPOs, namely $\mathcal{O}_1^{\leftrightarrows}$, $\mathcal{O}_2^{\leftrightarrows}$ and $\mathcal{O}_2^{\uparrow \downarrow}$. 
For LP fields (see Sec.~\ref{sec:LP_fields}) and high intensities, we have shown that only the RPOs of the family $\mathcal{O}_1$ persist. 
For ellipticities $\xi > 0$, the RPOs of the family $\mathcal{O}_{1}$ and $\mathcal{O}_{2}$ which counter-rotate with the IR field collapse at a critical intensity (see Figs.~\ref{fig:RPOs_OA_OF_py}c and~\ref{fig:stability_maps}), and therefore only the RPOs of the family $\mathcal{O}_{2}$ which co-rotate with the IR field persist. 
\par
We have shown that the location of RPOs in phase space indicate accurately the location of recolliding regions (see Fig.~\ref{fig:OA_OF_xpy_ellipticity}) which are sets of initial conditions in phase space which lead to recolliding trajectories under the evolution of Hamiltonian~\eqref{eq:main_hamiltonian}. 
This study has allowed us to determine the relations between the RPOs which have been identified in Refs.~\cite{Kamor2013, Mauger2014_JPB, Norman2015, Berman2015, Abanador2017} and used to predict and understand varying highly nonlinear phenomena.

\section*{Acknowledgments}
JD thanks Simon A. Berman, Cristel Chandre, Marc Jorba and Turgay Uzer for helpful discussions. JD thanks Fran\c{c}ois Mauger for sharing codes to generate Fig.~\ref{fig:illustration} using POV-Ray. The project leading to this research has received funding from the European Union's Horizon 2020 research and innovation program under the Marie Sk\l{}odowska-Curie grant agreement No.~734557. AJ has been supported by the Spanish grants PGC2018-100699-B-I00 (MCIU/AEI/FEDER, UE) and the Catalan grant 2017 SGR 1374.

\appendix
\section{Symmetries of Hamiltonian~\eqref{eq:main_hamiltonian} \label{app:symmetries}}
The equations of motion of Hamiltonian~\eqref{eq:main_hamiltonian} have the following symmetries:
\begin{itemize}
\item
If $\xi = 0$ (LP fields), the IR field~\eqref{eq:laser_field} has no components along the $\mathbf{e}_y$- and $\mathbf{e}_z$-axis. The equations of motion of Hamiltonian~\eqref{eq:main_hamiltonian} are invariant under rotations around the $\mathbf{e}_x$-axis, i.e., under the canonical transformation
\begin{subequations}
\begin{equation}
\label{eq:symmetry3}
(  \bar{\br}  ,  \bar{\bp} , \bar{t} ) = \left( \mathbf{R}_x (\vartheta) \br  , \mathbf{R}_x (\vartheta) \bp , t \right) ,
\end{equation}
for all $\vartheta$ and all time $t$, where $\mathbf{R}_x (\vartheta)$ is the rotation matrix around the $\mathbf{e}_x$-axis of angle $\vartheta$. As a consequence, the orbits in phase space are symmetric with respect to the rotation around $\mathbf{e}_x$. Another consequence is the invariance of the subspace $y = p_y = z = p_z = 0$. The electron belongs to this invariant subspace after ionization by the XUV pulse for $\Theta = 0$ and $\pi$.
\item
For all ellipticities, the IR field~\eqref{eq:laser_field} is such that $\mathbf{E}(t+T) = \mathbf{E}(t)$ and $\mathbf{E}(t+T/2) = - \mathbf{E}(t)$ for all time $t$. Hence, the equations of motion of Hamiltonian~\eqref{eq:main_hamiltonian} are invariant under the parity transformation, i.e., under the canonical transformation
\begin{equation}
\label{eq:symmetry1}
(\bar{\br} , \bar{\bp} , \bar{t}) = \left( - \br,- \bp , T/2 + t \right) .
\end{equation}
As a consequence, the orbits in phase space are symmetric with respect to the origin and the translation in time of half a laser cycle. 
\item
For all ellipticities, the IR field~\eqref{eq:laser_field} has no components along the $\mathbf{e}_z$-axis. Hence, the equations of motion of Hamiltonian~\eqref{eq:main_hamiltonian} are invariant under the parity transformation along the $\mathbf{e}_z$-axis, i.e., under the canonical transformation
\begin{equation}
\label{eq:symmetry2}
( \bar{\br} , \bar{\bp} , \bar{t} ) = \left( P_z (\br ) ,  P_z (\bp ) , t \right) ,
\end{equation}
where $P_z (x,y,z) = (x,y,-z)$. As a consequence, the orbits in phase space are symmetric with respect to the polarization plane $(\mathbf{e}_x , \mathbf{e}_y )$. Another consequence is the invariance of the subspace $z = p_z = 0$. For all values of XUV frequencies $\Omega$ and angles $\Theta$, the electron belongs to this invariant subspace, and therefore its dynamics is reduced to 2D ($d=2$).
\item
If $\xi = 1$ (CP fields), the IR field~\eqref{eq:laser_field} is such that $\mathbf{R}_z (\omega \tau) \mathbf{E}(t) = \mathbf{E}(t+\tau)$, where $\mathbf{R}_z (\omega \tau)$ is the rotation matrix around the $\mathbf{e}_z$-axis of angle $\omega \tau$. Hence, the equations of motion of Hamiltonian~\eqref{eq:main_hamiltonian} are invariant under rotations around the $\mathbf{e}_z$-axis and time translation, i.e., under the canonical transformation
\begin{equation}
\label{eq:symmetry4}
( \bar{\br} , \bar{\bp} , \bar{t} ) = \left( \mathbf{R}_z (\omega \tau) \br, \mathbf{R}_z (\omega \tau) \bp , t + \tau \right) .
\end{equation}
\end{subequations}
As a consequence, the orbits in phase space are symmetric with respect to the rotation of angle $\omega \tau$ around $\mathbf{e}_z$ and the translation in time $\tau$.
\end{itemize}

\section{Numerical calculation of periodic orbits and continuation methods \label{app:continuation_method}}
In this appendix, we provide the methods to compute the fixed point of a periodic orbit under a Poincar\'{e} map $\mathcal{P}$ for an evolving parameter $\mu$. The Poincar\'{e} map is given by 
\begin{equation}
\label{eq:app_Poincare_map}
\mathcal{P} \left( \mathbf{z} \right) = \boldsymbol{\varphi}_0^T \left( \mathbf{z} \right) ,
\end{equation}
where $\mathbf{z} = (\mathbf{r},\mathbf{p})$ and $\boldsymbol{\varphi}_0^T (\mathbf{z})$ is the flow of Hamiltonian~\eqref{eq:main_hamiltonian} for one period of the IR field $T$. The flow of the Hamiltonian is solution of Hamilton's equations $\dot{\mathbf{z}} = \lbrace \mathbf{z} , H \rbrace$, where $\lbrace \cdot , \cdot \rbrace$ denotes the canonical Poisson bracket. We denote $\mathbf{f} (\mathbf{z},t) = \lbrace \mathbf{z} , H \rbrace$ the Hamiltonian vector field, i.e., $\dot{\mathbf{z}} = \mathbf{f} (\mathbf{z},t)$. Periodic orbits of period $T$ (such as the RPOs $\mathcal{O}_1$, $\mathcal{O}_2$, $\mathcal{O}_1^{\leftrightarrows}$, $\mathcal{O}_2^{\leftrightarrows}$ and $\mathcal{O}_2^{\uparrow\downarrow}$) are fixed points under the Poincar\'{e} map~\eqref{eq:app_Poincare_map}. In what follows, fixed points of the Poincar\'{e} map~\eqref{eq:app_Poincare_map} are denoted $\mathbf{z}^{\star}$. The point in phase-space of coordinates $\mathbf{z}^{\star}$ is a fixed point under the Poincar\'{e} map~\eqref{eq:app_Poincare_map} if
\begin{equation}
\label{eq:app_zero_function_fixed_point}
\mathbf{F} (\mathbf{z}^{\star}) = \mathcal{P} (\mathbf{z}^{\star}) - \mathbf{z}^{\star} = \boldsymbol{0} .
\end{equation}
Hence, determining the fixed point of the Poincar\'{e} map~\eqref{eq:app_Poincare_map} consists of determining the zeros of the function~\eqref{eq:app_zero_function_fixed_point}. This is achieved using the Newton method. In order to do so, one needs to integrate Hamilton's equations and the tangent flow $\mathbb{J}_0^T (\mathbf{z}) = \partial \boldsymbol{\varphi}_0^T (\mathbf{z}) / \partial \mathbf{z}$ for one laser cycle. The differential equations governing the evolution of $\mathbb{J}_0^t (\mathbf{z})$ are given by
\begin{equation}
\label{eq:app_evolution_tangent_flow}
\dfrac{\mathrm{d}}{\mathrm{d}t} \mathbb{J}_0^t = \mathbb{A} \left( \boldsymbol{\varphi}_0^t(\mathbf{z})  , t\right) \mathbb{J}_0^t , \qquad \mathbb{A} (\mathbf{z},t) = \dfrac{\partial \mathbf{f} (\mathbf{z},t)}{\partial \mathbf{z}} .
\end{equation}
The initial conditions are $\mathbb{J}_0^0 (\mathbf{z}) = \mathbb{I}$, where $\mathbb{I}$ is the identity matrix. The linear stability of the periodic orbits, whose fixed point under the Poincar\'{e} map is $\mathbf{z}^{\star}$, corresponds to the eigenvalues of the monodromy matrix $\mathbb{M} (\mathbf{z}^{\star}) = \mathbb{J}_0^T (\mathbf{z}^{\star})$.
\par
Here, we describe the method to compute the fixed points $\mathbf{z}^{\star}$ of periodic orbits under the Poincar\'{e} map~\eqref{eq:app_Poincare_map} as a function of a parameter $\mu$. This method is used to follow the RPOs $\mathcal{O}_1$, $\mathcal{O}_1^{\leftrightarrows}$, $\mathcal{O}_2$, $\mathcal{O}_2^{\leftrightarrows}$ and $\mathcal{O}_2^{\uparrow\downarrow}$ as a function of the intensity of the IR field in Figs.~\ref{fig:recollision_probability},~\ref{fig:eigenvalues},~\ref{fig:RPOs_OA_OF_py}c and~\ref{fig:stability_maps} (i.e., with $\mu = E_0$). The Hamiltonian vector field and its flow depend explicitly on the parameter $\mu$, and as a consequence, they read $\mathbf{f} (\mathbf{z} , \mu , t)$ and $\boldsymbol{\varphi}_{0}^{t} (\mathbf{z},\mu)$, respectively [we note that the right-hand side of Eq.~\eqref{eq:app_Poincare_map} also depends explicitly on $\mu$]. 
The curve in the phase-parameter space for which $\mathbf{z}^{\star}$ is a fixed point of the Poincar\'{e} map~\eqref{eq:app_Poincare_map} for a given parameter $\mu$ is given by
\begin{equation}
\label{eq:app_g_zstar_mu_0}
g (\mathbf{z}^{\star} , \mu) = 0 .
\end{equation}
For instance, this curve corresponds to the union of the dashed and dotted curves in Fig.~\ref{fig:RPOs_OA_OF_py}c (with $\mu = E_0$).
The objective is to track the fixed points $\mathbf{z}^{\star}$ as a function of the parameter $\mu$ along this curve. We denote $\lbrace ( \mathbf{z}_{j}  , \mu_{j} ) \rbrace_j$ a set of points which belong to this curve, i.e., which fulfill the condition $g (\mathbf{z}_{j} , \mu_{j}) = 0$ for all $j$. We assume that a point $(\mathbf{z}_{0} , \mu_{0})$ is known. The intuitive way for tracking $\mathbf{z}^{\star}$ as a function of a parameter $\mu$ is to increase monotonically $\mu$ with a sufficiently small increment. However, we observe in Fig.~\ref{fig:RPOs_OA_OF_py} that the curve~\eqref{eq:app_g_zstar_mu_0} is not necessarily bijective with respect to $\mu$. Therefore, a more sophisticated method must be used.
\par
We use a continuation method which let the parameter $\mu$ free. We assume that a set of fixed points is known for a given set of parameters $(\mathbf{z}_j,\mu_j)$ with $j = 0 , \hdots , k-1$. The objective is to compute the fixed point $\mathbf{z}_k$ for a given parameter $\mu_k$. First, $\mathbf{z}_k$ is a fixed point for a given parameter $\mu_k$, and therefore it is a zero of Eq.~\eqref{eq:app_zero_function_fixed_point} which is
\begin{equation}
\label{eq:app_fixed_point_continuation_method}
\mathbf{F} ( \mathbf{z}_k , \mu_k ) = \mathcal{P} ( \mathbf{z}_k, \mu_k ) - \mathbf{z}_k = \boldsymbol{0} .
\end{equation}
Second, we introduce an increment $\delta$ in order to control the distance between two successive points $(\mathbf{z}_{k-1} , \mu_{k-1} )$ and $(\mathbf{z}_{k} , \mu_{k})$. The second equation which fixes $(\mathbf{z}_{k},\mu_{k})$ is given by
\begin{equation}
\label{eq:app_increment_delta_contiuation_method}
F_{\delta} ( \mathbf{z}_{k}, \mu_{k} ) = \left| \mathbf{z}_{k} - \mathbf{z}_{k-1} \right|^2 + \left| \mu_{k} - \mu_{k-1} \right|^2 - \delta^2 = 0 .
\end{equation}
This equation prevents the point $(\mathbf{z}_{k} , \mu_{k})$ to collapse to $(\mathbf{z}_{k-1} , \mu_{k-1})$. To conclude, given a point $(\mathbf{z}_{k-1},\mu_{k-1})$, the point $(\mathbf{z}_{k},\mu_{k})$ is solution of Eqs.~\eqref{eq:app_fixed_point_continuation_method} and~\eqref{eq:app_increment_delta_contiuation_method}. For each sample points along the curve $g$, we use the Newton method to determine the zeros $(\mathbf{z}_{k} , \mu_{k})$ of the function
\begin{equation*}
\mathbf{F}_{\mu} ( \mathbf{z}_{k} , \mu_{k} ) = 
\begin{bmatrix}
\mathbf{F} \left( \mathbf{z}_{k} , \mu_{k} \right) \\
F_{\delta}  \left( \mathbf{z}_{k} , \mu_{k} \right) 
\end{bmatrix} .
\end{equation*}
If $\delta$ is too large, problems of convergence for the Newton method can be encountered. In practice, the value of $\delta$ changes along the curve with respect to the number of iterations needed for the Newton method to converge. 
In the Newton method, Hamilton's equations, the tangent flow~\eqref{eq:app_evolution_tangent_flow} and the tangent flow with respect to the parameter $\mu$, i.e., $\mathbf{B}_0^t (\mathbf{z} , \mu) = \partial \boldsymbol{\varphi}_0^t (\mathbf{z},\mu)/\partial \mu$ are integrated over one period of the IR field. The evolution of the tangent flow with respect to the parameter $\mu$ is given by
\begin{equation*}
\dfrac{\mathrm{d}}{\mathrm{d}t} \mathbf{B}_0^t = \mathbb{A}\left( \boldsymbol{\varphi}_0^t (\mathbf{z},\mu),t \right) \mathbf{B}_0^t + \dfrac{\partial}{\partial \mu} \mathbf{f}(\boldsymbol{\varphi}_0^t \left( \mathbf{z},\mu),\mu,t \right) .
\end{equation*}
More details can be found in Appendix of Ref.~\citep{Dubois_thesis}.

%\bibliographystyle{apsrev4-1}
%\bibliographystyle{plain}
%\bibliography{biblio}

\end{document}